\documentstyle[12pt]{article}

\newtheorem{THEOREM}{Theorem}[section]
\newenvironment{theorem}{\begin{THEOREM} \hspace{-.85em} {\bf :} }%
                        {\end{THEOREM}}
\newtheorem{LEMMA}[THEOREM]{Lemma}
\newenvironment{lemma}{\begin{LEMMA} \hspace{-.85em} {\bf :} }%
                      {\end{LEMMA}}
\newtheorem{COROLLARY}[THEOREM]{Corollary}
\newenvironment{corollary}{\begin{COROLLARY} \hspace{-.85em} {\bf :} }%
                          {\end{COROLLARY}}
\newtheorem{PROPOSITION}[THEOREM]{Proposition}
\newenvironment{proposition}{\begin{PROPOSITION} \hspace{-.85em} {\bf :} }%
                            {\end{PROPOSITION}}
\newtheorem{DEFINITION}[THEOREM]{Definition}
\newenvironment{definition}{\begin{DEFINITION} \hspace{-.85em} {\bf :} \rm}%
                            {\end{DEFINITION}}
\newtheorem{CLAIM}[THEOREM]{Claim}
\newenvironment{claim}{\begin{CLAIM} \hspace{-.85em} {\bf :} \rm}%
                            {\end{CLAIM}}
\newtheorem{EXAMPLE}[THEOREM]{Example}
\newenvironment{example}{\begin{EXAMPLE} \hspace{-.85em} {\bf :} \rm}%
                            {\end{EXAMPLE}}
\newtheorem{REMARK}[THEOREM]{Remark}
\newenvironment{remark}{\begin{REMARK} \hspace{-.85em} {\bf :} \rm}%
                            {\end{REMARK}}

\newcommand{\thm}{\begin{theorem}}
\newcommand{\lem}{\begin{lemma}}
\newcommand{\pro}{\begin{proposition}}
\newcommand{\dfn}{\begin{definition}}
\newcommand{\rem}{\begin{remark}}
\newcommand{\xam}{\begin{example}}
\newcommand{\cor}{\begin{corollary}}
\newcommand{\prf}{\noindent{\bf Proof:} }
\newcommand{\ethm}{\end{theorem}}
\newcommand{\elem}{\end{lemma}}
\newcommand{\epro}{\end{proposition}}
\newcommand{\edfn}{\bbox\end{definition}}
\newcommand{\erem}{\bbox\end{remark}}
\newcommand{\exam}{\bbox\end{example}}
\newcommand{\ecor}{\end{corollary}}
\newcommand{\eprf}{\bbox\vspace{0.1in}}
\newcommand{\beqn}{\begin{equation}}
\newcommand{\eeqn}{\end{equation}}

\newcommand{\bbox}{\vrule height7pt width4pt depth1pt}

\newcommand{\clm}{\begin{claim}}
\newcommand{\eclm}{\end{claim}}

\newcommand{\sat}{\models}

\newcommand{\rimp}{\Rightarrow}

\newcommand{\bor}{\bigvee}

\newcommand{\union}{\cup}

\renewcommand{\phi}{\varphi}

\newcommand{\A}{{\cal A}}

\renewcommand{\P}{{\cal P}}

\newcommand{\R}{{\cal R}}

\newcommand{\W}{{\cal W}}

\newcommand{\<}{\langle}
\renewcommand{\>}{\rangle}

\newcommand{\eg}{e.g.,~}

\newcommand{\ol}{\setlength{\itemsep}{0pt}\begin{enumerate}}
\newcommand{\eol}{\end{enumerate}\setlength{\itemsep}{-\parsep}}
\newcommand{\ul}{\setlength{\itemsep}{0pt}\begin{itemize}}
\newcommand{\dl}{\setlength{\itemsep}{0pt}\begin{description}}
\newcommand{\edl}{\end{description}\setlength{\itemsep}{-\parsep}}
\newcommand{\eul}{\end{itemize}\setlength{\itemsep}{-\parsep}}

\newcommand{\commentout}[1]{}

\newcommand{\bi}{\begin{itemize}}
\newcommand{\ei}{\end{itemize}}
\newcommand{\be}{\begin{enumerate}}
\newcommand{\ee}{\end{enumerate}}

\newenvironment{oldthm}[1]{\par\noindent{\bf Theorem #1:} \em \noindent}{\par}
\newenvironment{oldlem}[1]{\par\noindent{\bf Lemma #1:} \em \noindent}{\par}
\newenvironment{oldcor}[1]{\par\noindent{\bf Corollary #1:} \em \noindent}{\par}
\newenvironment{oldpro}[1]{\par\noindent{\bf Proposition #1:} \em \noindent}{\par}
\newcommand{\othm}[1]{\begin{oldthm}{\ref{#1}}}
\newcommand{\eothm}{\end{oldthm} \medskip}
\newcommand{\olem}[1]{\begin{oldlem}{\ref{#1}}}
\newcommand{\eolem}{\end{oldlem} \medskip}
\newcommand{\ocor}[1]{\begin{oldcor}{\ref{#1}}}
\newcommand{\eocor}{\end{oldcor} \medskip}
\newcommand{\opro}[1]{\begin{oldpro}{\ref{#1}}}
\newcommand{\eopro}{\end{oldpro} \medskip}

\newcommand{\bxor}[1]{\dot{\bor}}

\renewcommand{\S}{{\cal S}}

\begin{document}

\renewcommand{\sc}{\sf}
\newcommand{\readfd}[2]{{\it suspect}_{#1}(#2)}
\newcommand{\readpfd}[2]{{\it suspect}'_{#1}(#2)}
\newcommand{\readfdp}[2]{{\it suspect}$'$_{#1}(#2)}
\newcommand{\readfdpp}{{\it suspect}$'$}
\newcommand{\m}{{\sf msg}}

\newcommand{\proc}	{\mbox{{\sf Proc}}}
\newcommand{\letset}[1]	{\mbox{{\sf #1}}}
\newcommand{\sset}	{\mbox{{\sf S}}}
\newcommand{\owner}[1]  {\mbox{\small $p_{#1}$}}
\newcommand{\OA}        {\mbox{$p_{\alpha}$}}
\newcommand{\OB}        {\mbox{$p_{\beta}$}}
\newcommand{\oa}        {\mbox{\small{\raisebox{-.6ex}{$p_{\alpha}$}}}}
\newcommand{\ob}        {\mbox{\small{\raisebox{-.6ex}{$p_{\beta}$}}}}

\newcommand{\quit}[1]     	{\mbox{{\it crash}$_{#1}$}}
\newcommand{\faulty}[1]     	{\mbox{{\it faulty}$_{#1}$}}
\newcommand{\start}[1]		{\mbox{{\it start}$_{#1}$}}
\newcommand{\recv}[3]           {\mbox{{\it recv}$_{#1}(#2, #3)$}}
\newcommand{\send}[3]           {\mbox{{\it send}$_{#1}(#2, #3)$}}
\newcommand{\mcast}[3]          {\mbox{{\it Mcast}$_{#1}(#2, #3)$}}
\newcommand{\learn}[2]		{\mbox{{\it learn}$_{#1}(#2)$}}
\newcommand{\try}[2]            {\mbox{{\it try}$_{#1}(#2)$}}
\newcommand{\susp}[2]		{\mbox{{\it suspect}$_{#1}(#2)$}}

\newcommand{\hist}[1]           {\mbox{$h_{#1}$}}
\newcommand{\hprime}[1]         {\mbox{$h'_{#1}$}}
\newcommand{\hbf}               {\mbox{$\rightarrow$}}
\newcommand{\hbfii}             {\mbox{$\stackrel{1}{\rightarrow}$}}
\newcommand{\cuteq}             {\mbox{$(h_1, \ldots, h_n)$}}
\newcommand{\cuteqprime}        {\mbox{$(h'_1, \ldots, h'_n)$}}
\newcommand{\pcuteq}            {\mbox{$(h_1, \ldots, h_p, \ldots, h_n)$}}
\newcommand{\pcuteqprime}       {\mbox{$(h'_1, \ldots, h'_p, \ldots, h'_n)$}}

\newcommand{\act}[2]		{\mbox{{\it do}$_{#1}(#2)$}}
\newcommand{\init}[2]		{\mbox{{\it init}$_{#1}(#2)$}}
\newcommand{\seq}[2]   		{\mbox{$\Sigma(#1,#2)$}}
\newcommand{\SigAB}   		{\mbox{$\Sigma(\alpha,\beta)$}}
\newcommand{\Term}[1]  		{\mbox{$\tau(#1)$}}
\newcommand{\TA}   		{\mbox{$\tau(\alpha)$}}



\newenvironment{proof}
    {{\sc Proof}}{  }

\newenvironment{proofsketch}
    {{\sc Proof Sketch}}{  }

\newcommand{\myenddef}{\rule{2mm}{3mm}}

\newtheorem{define}{Definition}

\newcommand{\gobble}[1]         {}
\newcommand{\AR}[1]             {{\bf ALETA:} #1 $\box$}
\newcommand{\ED}[1]             {{\bf ?`?`?` } #1 {\bf ???}}


\newenvironment{pgm}{
  \begin{tabbing}
  xxx \= xxx \= xxx \= xxx \= xxx \= xxx \= xxx \= xxx \= xxx \kill}{
  \end{tabbing}}

\newcommand{\ccomment}[1]{\mbox{/* #1 */}}


\newcommand{\never}             {\mbox{$\Box\neg$}}
\newcommand{\kk}[1]             {\mbox{{\large $K$}$_{#1}$}}
\newcommand{\ek}[1] {\mbox{{\large $E$}$_{\small{\raisebox{-.8ex}{#1}}}$}}
\newcommand{\ekk}[2]{\mbox{{\large
$E^{\small{#2}}_{\small{\raisebox{-.8ex}{#1}}}$}}}
\newcommand{\ck}[1] {\mbox{{\large $C$}$_{\small{\raisebox{-.8ex}{#1}}}$}}
\newcommand{\dk}[1] {\mbox{{\large $D$}$_{\small{\raisebox{-.8ex}{#1}}}$}}


\newcommand{\ar}[3]             {\mbox{{\sc recv}$_{#1}(#2, #3)$}}
\newcommand{\ess}[3]            {\mbox{{\sc send}$_{#1}(#2, #3)$}}
\newcommand{\ars}[3]            {\mbox{{\sc recv}$^{*}_{#1}(#2, #3)$}}
\newcommand{\suspect}[2]     	{\mbox{{\sc suspect}$_{#1}(#2)$}}

\newcommand{\actfla}[2]         {\mbox{{\sc do}$_{#1}(#2)$}}
\newcommand{\exe}[2]            {\mbox{{\sc exempt}$(#1,#2)$}}
\newcommand{\dexe}[2]           {\mbox{$\Diamond${\sc exempt}$(#1,#2)$}}
\newcommand{\quitfla}[1]        {\mbox{{\sc crash}$(#1)$}}
\newcommand{\alive}[1]          {\mbox{$\neg${\sc crash}$(#1)$}}
\newcommand{\dead}[1]           {\mbox{{\sc dis'd}$(#1)$}}
\newcommand{\simc}[1]           {\mbox{{\sc simCrash}$(#1)$}}
\newcommand{\ddead}[1]          {\mbox{$\Diamond\dead{#1}$}}
\newcommand{\dsim}[1]           {\mbox{$\Diamond\simc{#1}$}}
\newcommand{\aba}               {\mbox{{\sc try}$(\alpha)$}}
\newcommand{\abb}               {\mbox{{\sc try}$(\beta)$}}
\newcommand{\local}[2]          {\mbox{{\sc local}$_{#1}(#2)$}}

\newcommand{\shun}[2]           {\mbox{{\sc shun}$_{#1}(#2)$}}
\newcommand{\sshun}[1]          {\mbox{$\diamond$shun$(#1)$}}

\newcommand{\ifff}              {\mbox{$\Leftrightarrow$}}

\newcommand{\Pfd}{\mbox{${\cal P}$}}
\newcommand{\dP}{\mbox{$\diamond{\cal P}$}}
\newcommand{\Qfd}{\mbox{${\cal Q}$}}
\newcommand{\dQ}{\mbox{$\diamond{\cal Q}$}}
\newcommand{\Wfd}{\mbox{${\cal W}$}}
\newcommand{\dW}{\mbox{$\diamond{\cal W}$}}
\newcommand{\Wp}{\mbox{${\cal W}^{+}$}}
\newcommand{\Sfd}{\mbox{${\cal S}$}}
\newcommand{\dS}{\mbox{$\diamond{\cal S}$}}

\newcommand{\initfla}[2]{\mbox{{\sc init}$_{#1}(#2)$}}
\renewcommand{\dk}{D}
\renewcommand{\proc}{{\sf Proc}}
\newcommand{\citeyear}{\cite}
\newcommand{\psuspects}[1]{\mbox{{\sf Suspects}$_{#1}$}}

\begin{titlepage}
\title{A Knowledge-Theoretic Analysis of Uniform Distributed
Coordination and Failure Detectors%
\thanks{A preliminary version of this paper appeared in the 18th ACM
Symposium on Principles of Distributed Computing, 1999, pp.~73--82.}}
\author{Joseph Y.\ Halpern%
\thanks{This work was supported in part by NSF under
grants IRI-96-25901, by ONR under grant N00014-02-1-0455,
by the DoD Multidisciplinary University Research
Initiative (MURI) program administered by the ONR under
grants N00014-97-0505 and N00014-01-1-0795, and by a Guggenheim and a
Fulbright Fellowship. Sabbatical support from CWI and the Hebrew
University of Jerusalem is also gratefully acknowledged.
}\\
Cornell University\\
Dept. of Computer Science\\
Ithaca, NY 14853\\
\verb=halpern@cs.cornell.edu=\\
\verb=http://www.cs.cornell.edu/home/halpern=\\
\and
Aleta Ricciardi\\
University of Texas at Austin \\
Dept. of Electrical and Computer Engineering\\
Austin, TX 78712\\
\verb=aleta.ricciardi@ece.utexas.edu= \\
\verb=http://www.nile.utexas.edu/~aleta=
}
\date{ }
\setcounter{page}{0}
\maketitle
\thispagestyle{empty}

\begin{abstract}
It is shown that, in a precise sense, if there is no bound on the number
of faulty processes in a system with unreliable but fair
communication, Uniform Distributed Coordination (UDC) can be
attained if and only if a system has {\em perfect\/} failure detectors.
This result is generalized to the case where there is a bound $t$
on the number of faulty processes.  It is shown that a certain type of
generalized failure detector is necessary and sufficient for achieving
UDC in a context with at most $t$ faulty processes.  Reasoning about
processes' knowledge as to which other processes are
faulty plays a key role in the analysis.
\end{abstract}
\end{titlepage}

\section{Introduction}
\label{SecIntro}

Periodically coordinating specific actions among a group of processes is
fundamental to 
solving
most distributed computing problems, and especially to
replication schemes that achieve fault tolerance.  Unfortunately, as is
well known, it is impossible to achieve coordination in an asynchronous
setting even if there can be only one faulty process \cite{FLP}.
This is true even if communication is reliable.
As a result, there has been a great deal of interest recently in systems
with {\em failure detectors\/} \cite{CT96}, oracles that provide
suspicions as to which processes in the system are faulty.  This interest is
heightened by results of Chandra, Hadzilacos, and Toueg
\citeyear{CT96,CHT96} showing that consensus can 
be achieved with relatively unreliable failure detectors

Here we consider what kind of failure detectors are necessary to attain
{\em Uniform Distributed Coordination\/} (UDC) \cite{GT-WDAG89}.  We
have UDC of action
$\alpha$ if, whenever some process (correct or not) performs $\alpha$,
then so do all the 
correct processes.  
There are two features that distinguish UDC from consensus.
First, if a process that initiates an action is later found to be
faulty, in the UDC setting, all the processes must still perform the
action.  On the other hand, in the case of consensus, the nonfaulty
processes can agree not to perform the action.  This property of UDC is
particularly important in practice.  Consider for example, a group of
processes implementing
fault-tolerant service; actions are executed on behalf of clients and
change the state of the service (for example, allocating a scarce
resource). In the UDC setting, the service cannot repudiate an action
should the member eventually be deemed faulty, as could be the case in
consensus. With UDC, the service is required to make that action part of the
service's communal history. From the client's point of view, the eventual
designation of a group member as faulty is irrelevant; indeed, one goal of
using replication to implement a service is to mask failures from clients.
A second difference between UDC and consensus is that,
in consensus, processes must typically choose exactly one out of
two actions (``attack'' or ``retreat''; or, ``decide 0'' or ``decide
1'').  On the other hand, in UDC,
there is no choice to be made; that is, UDC has no requirement that if
action $\alpha$ is ever taken, then of necessity, action $\beta$ is
never taken.  
Thus, UDC suffices whenever actions to be taken by a group can be partitioned
into non-conflicting subsets; 
it requires consenses to decide which of a conflicting set of actions to
take.  

If we have reliable communication,
then it is easy to see that we can attain UDC no matter how many
processes may fail.  Thus, in this setting, UDC is strictly easier than
consensus.  
Intuitively, consensus requires all the correct processes to agree on a
particular action.  For example, they must all agree to attack or all
agree to retreat (but cannot do both).  With UDC, if one process
attacks, all the correct processes must attack, and if one retreats, all must
retreat.  But it is perfectly consistent with UDC for the correct
processes 
both to attack and to retreat.

If communication is unreliable but fair, then we show that we 
can attain UDC 
even if there is no bound on the number of process failures 
(that is, even if there are runs in which all processes may fail)
in the presence of {\em weak\/} failure
detectors, which have the property that eventually 
each faulty process is permanently suspected by at least one correct 
process ({\em weak completeness\/})
and at least one correct
process is never suspected ({\em weak accuracy}).  Chandra and Toueg
\citeyear{CT96} showed that consensus with an arbitrary number of failures
is also achievable using weak failure detectors.  They considered a
setting with reliable communication, but their results apply with
essentially no change to a setting where communication is unreliable but
fair in an appropriate sense.

Chandra and Toueg
observed that by having processes
communicate their suspicions, a weak failure detector can be converted to a
{\em strong\/} failure detector, which satisfies weak accuracy and {\em
strong\/} completeness (all correct processes eventually permanently suspect
every faulty process).
We further show that, under an assumption about the independence of
process failures, 
in systems with no bound on the number of faulty 
processes, strong failure detectors are equivalent to {\em perfect\/}
failure detectors, which satisfy strong completeness and {\em
strong accuracy\/}---%
no process is suspected until it crashes.  
(Indeed, under the same conditions, weak failure detectors are
equivalent to perfect failure detectors.)  

These results tell us that, if there is no bound on
failures, then we can attain UDC using what are effectively
equivalent to perfect failure detectors.   Are perfect failure detectors
really necessary? We show that in a
precise sense they are.  Under quite minimal assumptions, 
perfect failure detectors can be implemented in a system that
attains UDC with no bounds on the number of failures.%
\footnote{We remark that our notion of ``implement'' is stronger than
the notion of reduction used by Chandra, Hadzilacos, and Toueg \citeyear{CT96,CHT96}; see
Section~\ref{SecKA}.}
 It is interesting to note
that Schiper and Sandoz' {\em Uniform Reliable
Multicast}~\citeyear{SchiperSandoz93} is a special case of UDC where the
only action of interest is reliable message delivery.  Schiper and
Sandoz implement Uniform Reliable Multicast by using the Isis virtual
synchrony model~\cite{BJ87}, which simulates perfect failure detection.
Our results support their need to implement it in this way.

What happens if there is a bound on the number of faulty processes?
Gopal and Toueg \citeyear{GT-WDAG89}
show that UDC is achievable with no failure detectors 
in systems where fewer than half the processes can fail.
Here we generalize these
results, providing, for each value of $t$, a generalized failure
detector that we can show is necessary and sufficient to attain UDC if
there are at most $t$ failures.
The generalized failure detector we consider reports suspicions of the
form ``at least $k$ processes in a set $S$ of processes are faulty''
(although it does not specify which $k$ are the faulty ones).  Such
generalized failure detectors may be appropriate when the system can be
viewed as consisting of a number of components, and all we can say is
that some process in a component is faulty, without being able to say
which one it is.

The rest of this paper is organized as follows.  In
Section~\ref{SecBackground}, we provide the necessary background,
reviewing the formal model, failure detectors, the formal language, and
the definition of UDC.  In Section~\ref{SecKA}, we present our analysis
in the case that there is no bound on the number of faulty processes.
Our proof techniques may be of
independent interest, since 
they make nontrivial  use of 
the knowledge-theoretic tools of Fagin et al.~\citeyear{FHMV}.
Reasoning about the
knowledge of the processes in the system---particularly, their knowledge
of which other agents are faulty---plays a key role in the analysis.
In Section~\ref{SecGen}, we extend this analysis to the case where there 
is a known bound $t$ on the number of faulty processes; 
we also introduce our generalized failure detectors.
We conclude in Section~\ref{sec:conclusions} 
with a discussion of the results and a comparison of our results to
results of
Aguilera, Toueg, and Deianov \citeyear{ATD99} who, in response to the
conference version of this paper \cite{HR99}, provided an 
alternative
characterization of the type of failure detectors needed to attain UDC.
Proofs are relegated
to the Appendix.

\section{Background}
In this section, we briefly discuss the formal model (and, in particular,
our assumptions about message delivery), failure detectors, the formal
language that we use for expressing coordination, which includes
operators for knowledge and time, and the notion of UDC.

\label{SecBackground}

\subsection{The Model}
We adopt the familiar model of an asynchronous distributed system.  We
assume that there is a fixed finite
set $\proc = \{p_1, \ldots, p_n\}$ of processes with no shared
global clock. These processes
communicate with one another by passing messages over a completely
connected network of channels.
Processes fail by crashing and do not recover, but otherwise follow their
assigned protocols.
Channels are not reliable.  A message that is sent is not necessarily
received and, even if it is received, there is no 
upper
bound on
message transmission delay.  However, channels do not
corrupt messages (so that every message received is one that was
actually sent) and they are
{\em fair}, in the sense that if the same message is sent from $p$ to
$q$ infinitely often and $q$ does not crash, then the message is
eventually received 
infinitely often
by $q$.
Processes and the environment (or nature) execute {\em actions};
corresponding to each action is an {\em event} (intuitively, the event
of that action occurring).  
We assume that the events that take place at a particular process are
totally ordered, and are recorded in that process's {\em history}.
The events recorded in $p$'s history include
communication events 
of the form 
\send{p}{q}{\m} ($p$ sends message $\m$ to
$q$) and
\recv{p}{q}{\m} ($p$ receives $\m$ from $q$);
internal events, which include events of the form $\act{p}{\alpha}$
($p$ executes action $\alpha$)
and
$\init{p}{\alpha}$
($p$ initiates $\alpha$;
see Section~\ref{SecUDC}); 
the special event \quit{p},
which models the failure of $p$;
and {\em failure-detector events}, which are discussed in
Section~\ref{SecFDs}.  

A {\em history} for process $p$, denoted $h_p$,
is a sequence of events corresponding to actions performed by process
$p$.
A {\em cut\/} is
a tuple of finite process histories, one for each $p\in\proc$.
A {\em run\/} is a function from time (which we take to range over
the natural numbers, for simplicity) to cuts.  If $r$ is a run, we use
$r_p(m)$ to denote $p$'s history in the cut $r(m)$.
A pair $(r,m)$ consisting of a run $r$ and a time $m$ is called a {\em
point}.  We write $(r,m) \sim_p (r',m')$ if $r_p(m) = r'_p(m')$.
We say that a run $r'$ {\em extends\/} a point $(r,m)$ if $r'(m') =
r(m')$ for all $m' \le m$.  Thus, $r'$ extends $(r,m)$ if $r$ and
$r'$ have the same prefix up to time $m$.  
Process $q$ is faulty in run $r$ iff $\quit{q}$ is in $q$'s history.
$F(r)$ denotes the faulty processes in run $r$.

We assume that a run $r$ satisfies the following.
\begin{itemize}
\item[R1.] $r(0) = (\<\,\>,\ldots, \<\,\>)$ (that is, at time 0, each
process's history is empty).
\item[R2.] $r_p(m+1) = r_p(m)$ or $r_p(m+1)$ is the result of appending
one event to $r_p(m)$.
\item[R3.] If \recv{q}{p}{\m} is in $r_q(m)$, then the
corresponding send event \send{p}{q}{\m} is in $r_p(m)$.
\item[R4.] If the event \quit{p} is in $r_p(m)$, then 
it is the last event in $r_p(m)$.
\item[R5.] If the number of occurrences of
\send{p}{q}{\m} in $r_p(m)$ grows unboundedly as $m$ increases,
then either the event \quit{q} appears in
$r_q(m)$ for some $m$ or the number of occurrences of \recv{q}{p}{\m} in
$r_q(m)$ grows unboundedly as $m$ increases.  
(Informally, if in run $r$ process $p$ sends $\m$ infinitely often to
$q$, then either $q$ crashes or $q$ receives $\m$ infinitely often.)
\end{itemize}
When we consider failure detectors, we add
further conditions to runs.

A {\em system\/} is a set of runs.
Systems are typically generated by {\em protocols} executed in a certain {\em
context}.  Formally, a {\em protocol
for process $p$\/} is
a function from finite histories to actions.
A {\em joint protocol\/} is a tuple $(P_1,\ldots, P_n)$ consisting of a
protocol
for each process in $\proc$.   A run $r$ is consistent with a
joint protocol
$P$ if, for all times $m_1$, if $r_p(m_1+1) = r_p(m_1) \cdot e$ and
$e$ is an event corresponding to a protocol action, then $e$ is in fact
the event 
corresponding
to the action $P_i(r_p(m_1))$.  A {\em context\/} for us is
simply a bound on the number of processes that can fail 
(if there is such a bound),
a specification of properties of failure detectors
(see Section~\ref{SecFDs},
and a specification of communication properties (whether communication
is reliable, fair, etc.).
Fagin et al.~\cite{FHMV,FHMV94} give a 
more general
definition of context, but this suffices for our purposes).
In a given context,
a joint
protocol generates the system consisting of all the runs satisfying
R1--R5 and the constraints of the context that are consistent with the
protocol. 
We say that a 
joint
protocol has a certain property 
in a given context if the system it generates in that context has that
property. 
Note that, because all runs in the systems we consider are assumed
satisfy R5, we are restricting in this paper to systems where
communication is fair, although possibly unreliable.

\subsection{Failure Detectors}
\label{SecFDs}

Informally, a {\em failure detector\/}~\cite{CT96} is a per-process
oracle that emits suspicions regarding other processes' faultiness.
The fact that a process
$q$ is  suspected by process $p$'s failure detector does not mean that
$q$ is in fact faulty.  
Various failure detectors can be defined by imposing conditions 
on the accuracy and completeness of suspicions. 
Chandra and Toueg \citeyear{CT96} model failure detectors by assuming a
function $H$ such that $H(p,t)$ describes the suspicions of $p$'s failure
detector at time $t$.  Chandra and Toueg then assume that processes
explicitly query their failure detectors to ``learn'' those suspicions.
Our approach is slightly more general.
We model the act of $p$
getting a report $x$ from 
its failure detector by the event $\readfd{p}{x}$.  
A process $p$ could ``get
a report from its failure detector'' either because it explicitly reads
it (as Chandra and Toueg assume) or because the failure
detector automatically emits a suspicion.   A {\em standard report\/} is
one of the form ``the processes in $S$ are faulty'', which we model by
the report $\readfd{p}{S}$.   A {\em standard failure detector\/} is one
whose reports are standard.  In a system with standard failure detectors, 
at each point $p$, define $\psuspects{p}(r,m) = S$ if and only if
$\readfd{p}{S}$ is the most recent failure-detector event in $r_p(m)$.
(If there have not been any reports by time $m$ in $r$,
$\psuspects{p}(r,m) = \emptyset$.)
We will shortly generalize the definition of $\psuspects{p}(r,m)$ so
that it applies in the presence of (some)
nonstandard failure detectors.
The differences between our way of modeling failure detectors 
and the Chandra-Toueg approach
are mainly cosmetic.  In the Chandra-Toueg approach, what we are
calling a run
consists of the actions performed by the processes 
(including reading the failure detector) 
and a special tape or ``oracle'' that
describes the responses when the failure detector is read.
We have used our approach so as to be
able to capture all the behavior of the system in terms
of histories, without invoking
any extra structure (such as extra tapes).  It is easy to translate
from runs in the Chandra-Toueg framework to runs in our framework, and
vice versa.
Given a run in the Chandra-Toueg framework, the corresponding run in our
framework uses the event
``$\readfd{p}{x}$''
indicates both that $p$ read its special
tape and that the response was $x$. 
Conversely, given a run 
in our framework, 
the corresponding run in the Chandra-Toueg framework has $p$ query its
failure detector and receive response $x$ at exactly the points where
the event $\readfd{p}{x}$ appears in its history.

Although there is a one-to-one mapping between runs in our framework
and runs in the Chandra-Toueg framework, the systems (i.e., sets of
runs) that we allow
are slightly more general than the systems they consider.
Chandra and Toueg essentially consider only systems that are
a cross-product of the set of possible special tapes and the set
of possible actions performed by the processes.  That is, no correlation
is allowed between the behavior of the processes and the behavior of the
failure detector.  We do allow correlation, and thus can consider types
of failure detectors that Chandra and Toueg cannot (see below).  
However, it would be easy to extend the Chandra-Toueg framework to allow
such correlation.

Consider the following properties of 
standard
failure detectors
(the first four are
also used by Chandra and Toueg):
\begin{description}

\item[{\em Strong Accuracy:}]
No process is suspected before it crashes.  Formally,
for all processes $p$ and $q$  and times $m$,
if $q \in \psuspects{p}(r,m)$, then 
$\quit{q}$ is in $r_q(m)$.
\item[{\em Weak Accuracy:}]
If there is a correct process, then
some correct process is never suspected.  Formally, if $F(r) \ne\proc$
then there is some $q \notin F(r)$ such that,
for all processes $p$ and times $m$,
$q \notin \psuspects{p}(r,m)$.%

\item[{\em Strong Completeness:}] All faulty processes are eventually
permanently suspected by all correct processes.  Formally, if $q \in
F(r)$ and $p \notin F(r)$,
then there is a time $m$ such that for all $m'\geq m$, $q\in
\psuspects{p}(r,m')$.

\item[{\em Weak Completeness:}] Each faulty process is eventually
permanently suspected by some correct process.
Formally, if $q \in F(r)$ 
and $F(r) \ne \proc$,
then there exists some $p \notin F(r)$ and
a time $m$ such that, for all $m'\geq m$, $q\in
\psuspects{p}(r,m')$.%
\footnote{Chandra and Toueg do not require that
$F(r) \ne \proc$ in their definition of weak accuracy
or weak completeness,
since they
assume
that there always is
at least one correct process.  We have added it here 
since we allow runs where all processes fail.}

\item[{\em Impermanent Strong Completeness:}] All faulty processes are
eventually suspected (but not necessarily permanently) by all correct
processes.  Formally, if $q \in F(r)$ and $p \notin F(r)$, then there is
some time $m$ such that $q \in\psuspects{p}(r,m)$.

\item[{\em Impermanent Weak Completeness:}] 
Each faulty process is
eventually suspected (but not necessarily permanently) by some correct
process.  Formally, if $q \in F(r)$
and $F(r) \ne \proc$,
then there 
is some $p \notin F(r)$
and time $m$
such that 
$q \in\psuspects{p}(r,m)$.

\end{description}
\noindent A system $\R$ is said to satisfy a given property of failure
detectors (e.g., weak completeness or impermanent strong completeness)
if the failure detectors in every run of $\R$ satisfy the property.

We remark that impermanent strong and weak completeness cannot 
meaningfully be captured in the
Chandra-Toueg framework because, as we mentioned earlier, Chandra and
Toueg do not allow correlation between the behavior of processes and the
behavior of the failure detector.
The only way that a
special tape can guarantee impermanent strong completeness is to
ensure that eventually, whenever the tape is constructed, it will report
a failure.  (Otherwise it might be consulted only at times when it does
not report a failure.)  Impermanent strong completeness requires a
correlation between the special tape and the actions of the
processes of a sort not allowed by Chandra and Toueg.

Chandra and Toueg define a {\em perfect failure detector\/}
as one that satisfies strong completeness and strong accuracy,
a {\em strong failure detector\/}
as one that satisfies strong
completeness and weak accuracy, and a {\em weak failure detector\/} 
as one that satisfies weak completeness and weak accuracy.  We define 
an
{\em impermanent-strong
failure detector\/} 
as one that satisfies impermanent
strong completeness and weak accuracy and 
an
{\em impermanent-weak
failure detector\/}
as one that satisfies impermanent weak completeness
and weak accuracy.   
The definitions above have focused on standard failure detectors, whose
reports have the form ``the processes in $S$ are faulty''.  However,
other types of reports can also be used, as long as they 
can be viewed as saying
that the processes in some set $S$ are faulty.  For example, a
report of the form ``the processes in $\proc - S$ are correct'' can be
clearly viewed as saying the processes in $S$ are faulty.  To make this
precise, 
we say that a failure detector is {\em $g$-standard\/} if $g$ is a 
function mapping the reports of the failure detector to subsets of $\proc$.
Thus, the failure detector that reports that the processes in $\proc -
S$ are correct 
(such failure detectors are used in \cite{ATD99}, for example)
is $g$-standard, where $g(\mbox{``the processes in $\proc
- S$ are correct''}) = S$.
If $p$ has a $g$-standard failure detector, define $\psuspects{p}(r,m) =
S$ if and only if $\readfd{p}{x}$ is the most recent failure-detector
event in $r_p(m)$ and $g(x) = S$.  
Notions of strong accuracy, weak
accuracy, and so on now apply to $g$-standard failure detectors with no
change in the definition.  
Although we consider only standard failure detectors in this paper, 
all of our results apply to $g$-standard failure detectors as well.

Chandra and Toueg show that a failure detector satisfying weak
completeness 
can be converted
to one satisfying strong
completeness, while still preserving accuracy properties.
Roughly speaking, all processes just communicate and tell each other
about the suspicions reported by their original failure detectors; their
modified failure detector reports all the suspicions they hear about.
The same construction can be used to convert a failure detector
satisfying 
weak impermanent completeness
to one satisfying 
strong impermanent completeness.  
We need to be a little careful in making precise in our framework the
notion of converting one type of failure detector to another.  In the
simplest case, given a system $\R$, it is 
simply a question of considering a system $\R'$ where each run $r
\in \R$ is replaced by a run $r' \in \R'$ such that each occurrence of 
an event of the form $\readfd{p}{x}$ is replaced by a different 
failure-detector event $\readfd{p}{x'}$, reflecting the modified failure
detector.  However, if the conversion also involves additional
communication (as in the conversion from weak completeness to strong
completeness), the naive replacement of failure-detector events does not
suffice.  Nevertheless, the basic notion of conversion remains the same.
Namely, we assume that there is some function $f$ mapping runs to runs
such that all the events in $r$ (except possibly the failure-detector
events) appear as events in $f(r)$, and appear in the same order in 
$r$ and $f(r)$.  However, $f(r)$ may have additional events, including
additional communication between processes and new failure-detector
events.  
These new failure-detector events are the ones that we consider in
determining whether $\R'$ has failure detectors that satisfy properties
such as  strong completeness.   
In Section~\ref{SecKA}, we use a particular instance of this conversion
process to show how systems that allow solutions to UDC can be used to
implement failure detectors with certain properties.
For now, we leave it to the reader to check that Chandra and Toueg's
conversion from failure detectors satisfying weak completeness to ones
satisfying strong completeness can be implemented in this framework.
This gives us the following result.

\pro\label{convert} A system $\R$ with weak (resp.,
impermanent-weak) failure detectors can be converted to a system $\R'$
with strong (resp., impermanent-strong) failure detectors, while
preserving accuracy properties.
\epro

Note that we can trivially convert a
failure detector that satisfies
impermanent strong completeness to one that
satisfies strong completeness by
always outputting the list of all
previously suspected processes.  
For convenience, we state this as a separate proposition.
\pro\label{convert1} A system $\R$ with impermanent-strong failure 
detectors can be converted to a system $\R'$ with strong failure
detectors, while preserving accuracy properties.
\epro

As we show in Section~\ref{SecKA},
under a minimal assumption that should surely be satisfied in practice,
if there is no bound on the number of faults
(i.e., if there are runs where all processes may fail),
a failure detector that satisfies weak accuracy must also satisfy strong
accuracy.  Thus, if there is no bound on the number of
faults, then there is essentially
no difference between
impermanent-weak failure detectors and perfect
failure detectors.\footnote{In the notation of Chandra and Toueg,
impermanent-$\W \cong$ impermanent-$\S \cong \S \cong \P$ for $t =
n-1$  
or $t=n$
failures. }

\subsection{The Formal Language}
\label{SecLanguage}

Our language for reasoning about distributed coordination involves
time and knowledge.  The underlying notion of time is linear (so our
language extends linear time temporal logic).
We find it useful to be
able to reason about the past as well as the future.
Formally, we start with 
(application-dependent)
primitive propositions and close
under Boolean combinations, $\Box$,
and the epistemic operators $K_p$
 for each process $p$.
Following \cite{FHMV}, we define the truth of a formula
relative to a tuple $(\R,r,m)$ consisting of
a system $\R$, run $r \in \R$, and time $m$.
We write $(\R,r,m)\models
\varphi$ if the formula
$\varphi$ is true at the point $(r,m)$ in system $\R$.  Among the
primitive propositions in the language are
\ess{p}{q}{\m}, \ar{q}{p}{\m}, \quitfla{p}, \actfla{p}{\alpha}, and
\initfla{p}{\alpha}.
The truth of these primitive propositions is
determined by
the cut in the obvious way;  for example, \ess{p}{q}{\m} is true at
a cut precisely
when \send{p}{q}{\m} is an event in $p$'s history component of the cut.
$\Box \varphi$ holds at a point if $\phi$ holds from that point on in the
run. Thus, $(\R,r,m) \models
\Box\varphi$ if and only if
$(\R,r,m')\models \varphi$ for all $m' \ge m$.
As usual, we define $\Diamond\varphi = \neg \Box \neg \varphi$; thus,
$\Diamond$ is the dual of $\Box$.  It is easy to see that $(\R,r,m)
\models
\Diamond \varphi$ if $(\R,r,m') \models \varphi$ for some $m' \ge m$.
Finally, $K_p \varphi$ is true if $\varphi$ is true at all the points
that $p$ considers possible, given its current history.  Formally,
$(\R,r,m) \models K_p\varphi$ if and only if $(\R,r',m') \sat \phi$
for all points $(r',m') \sim_p (r,m)$
such that $r' \in \R$.
We say a
formula $\varphi$ is {\em valid in system $\R$},
denoted $\R \models \varphi$, if $(\R,r,m) \models \varphi$ for all
points $(r,m)$ in $\R$.

In our analysis, we make particular use of {\em local\/} and
{\em stable\/} formulas. A formula $\varphi$ is 
{\em local}
to process $p$ in system $\R$ if at every point
in $\R$, $p$ knows whether $\phi$ is true, that is, 
$\phi$ is local to $p$ in $\R$
if $K_p \varphi \lor K_p \neg
\varphi$ is valid in $\R$.  
All formulas describing a process's
local state, for example, \ess{p}{q}{\m}, \ar{q}{p}{\m}, \quitfla{p},
and \initfla{p}{\alpha}, are local
to that process.
It follows from standard properties of knowledge (see \cite{FHMV})
that formulas of the form
$K_p \varphi$ are also local to $p$, since
$K_p \bigl(K_p\varphi\bigr) \vee K_p (\neg K_p \varphi)$
is valid in every system.
A {\em stable\/} formula is one that, once true, remains true; that is,
$\phi$ is stable in system $\R$ if $\phi \rimp \Box \phi$ is valid in
$\R$.   All of \ess{p}{q}{\m}, \ar{q}{p}{\m}, \quitfla{p}, 
\initfla{p}{\alpha}, and $\Box \phi$ are stable.

\subsection{Distributed Coordination}
\label{SecUDC}

We are interested in modeling distributed coordination of certain
actions among the processes in $\proc$.
The actions may be allocating a resource, delivering multicast messages, or
committing a transaction; we are not concerned with the specifics.
We are also not concerned here with other requirements such as executing
actions in a particular order (\eg total-order multicast) or not executing
conflicting actions (\eg consensus).  We are interested only in the
eventual, distributed execution of these actions.

Formally, we assume that each process $p$ has a set $\A_p$ of {\em
coordination actions\/} that it can initiate.   We assume that the sets
$\A_p$ and $\A_q$ are disjoint for $p \ne q$.  (Think of the actions in
$\A_p$ as somehow being tagged by $p$.)  
The fact that an action $\alpha$ is in $\A_p$ does {\em not\/} mean that only
$p$ can perform $\alpha$.  However, it does mean that only $p$ can
initiate $\alpha$; no process can perform $\alpha$ unless $p$ initiates it.
We assume that for each action 
$\alpha \in \A_p$, there is a special action 
\initfla{p}{\alpha} of $p$ {\em initiating 
$\alpha$}.  The corresponding event $\init{p}{\alpha}$ can appear only in
$p$'s history, and can appear at most once in a run.  
Formally, for the rest of this paper, we consider only systems $\R$
where, for all points $(r,m)$ in $\R$ and all actions $\alpha \in \A_p$, the
event $\init{p}{\alpha}$ can appear only in $r_p(m)$ and can appear at
most once in $r_p(m)$.
Informally, a system satisfies
{\em Uniform Distributed Coordination\/} (UDC) of action
$\alpha$ 
if whenever any
$p' \in \proc$ executes $\alpha \in {\cal A}_p$, then
so eventually  does every correct
$q \in \proc$.
In addition,
no process performs $\alpha\in {\cal A}_p$ unless $p$ initiates it.
Intuitively, if $\init{p}{\alpha}$ appears in $p$'s history
and $p$ is nonfaulty in run $r$, then all the nonfaulty processes 
in $r$ should perform $\alpha$.  
Formally, UDC of $\alpha \in \A_p$ holds in a system
$\R$ if the following three
conditions hold:
\begin{itemize}
\item[DC1.]
$\R \sat \initfla{p}{\alpha} \rimp \Diamond ( \actfla{p}{\alpha} \vee
\quitfla{p} )$;
\item[DC2.] $\R \sat \bigwedge_{q_1,q_2 \in \proc} \Bigl(
\actfla{q_1}{\alpha} \Rightarrow 
\Diamond( \actfla{q_2}{\alpha} \vee
\quitfla{q_2}) \Bigr)$;
\item[DC3.]
$\R \sat
\bigwedge_{q_2 \in \proc} \Bigl( \actfla{q_2}{\alpha} \Rightarrow
\initfla{p}{\alpha} \Bigr)$.
\end{itemize}

{\em Non-Uniform Distributed Coordination\/} (nUDC) requires
coordination only if the process that performs $\alpha$ is correct.
Thus, nUDC of $\alpha$ holds in a system $\R$ if
DC1, DC3, and the following hold:
\begin{itemize}
\item[DC2$'$.]
$\R \sat \bigwedge_{q_1,q_2 \in \proc} \Bigl( \actfla{q_1}{\alpha} \Rightarrow
\Diamond( \actfla{q_2}{\alpha} \vee
\quitfla{q_2} \vee \quitfla{q_1}) \Bigr)$.
\end{itemize}

The next propositions show that, unlike UDC, nUDC is easy to attain, and
that reliable communication is significant for UDC.
(As we said earlier, all proofs are in the Appendix.)

\pro\label{nUDC} 
There is a protocol that attains nUDC without the use of failure detectors
in every context where communication is fair (although possibly
unreliable), even if there is no bound on the number of failures.
\epro

\pro\label{UDCok}  There is a protocol that attains UDC
without the use of failure detectors in every context 
where communication is reliable, even if there is no bound on the
number of failures. 
\epro

Propositions~\ref{nUDC} and \ref{UDCok}
distinguish UDC and nUDC from consensus.
Unlike consensus, both UDC and nUDC are
attainable in asynchronous systems with failures (although
UDC needs reliable communication); indeed, they are attainable 
without failure detectors
no matter how many
processes may fail. However, as we shall see in the next two sections,
things change when we consider UDC
in a context with unreliable communication.

\section{UDC With No Bound on Failures}
\label{SecKA}

We start by showing that UDC is achievable in a context with 
fair but
unreliable
communication, provided we have
impermanent-strong failure detectors.

\pro\label{UDC} There is a protocol that attains UDC in 
every context
with strong failure detectors,
even if there is
no bound on the number of failures.
\epro

In light of Proposition~\ref{convert} and~\ref{convert1}, the following
corollary is immediate.

\cor\label{UDC2} There is a protocol that attains UDC in 
every context with impermanent-weak failure detectors,
even if there is no bound on the number of failures.
\ecor

Chandra and Toueg~\citeyear{CT96} prove a result analogous to
Proposition~\ref{UDC} for consensus.  They show that consensus is
achievable in every context where there are strong failure detectors, at
most $n-1$ failures, and where communication is reliable.  Their
algorithm works without change even if we have only
impermanent-strong failure
detectors and allow $n$ failures.  Moreover, their algorithm can be
modified easily to deal with unreliable, but fair, communication.  Thus,
unlike UDC,
the reliability of communication
has no significant impact 
on the attainability of consensus
in these contexts.

We prove
in Theorem~\ref{UDCchar} below that under certain assumptions
about the context (which include the assumption 
that
there is no bound on the
number of failures
along with our usual implicit assumption that communication is fair,
although possibly unreliable),
if processes can perform UDC
then they can simulate perfect failure detectors.
It follows from Proposition~\ref{strong=perfect} 
below
that, under these
assumptions, strong failures detectors are equivalent to perfect
failure detectors.  Thus, we will be proving what is essentially 
a converse to Proposition~\ref{UDC}.
To prove this result, 
we need to  make precise 
the notion of ``simulating a perfect failure detector.'' 
``Simulating a perfect failure detector'' means that 
we can convert a system $\R$ to a system $\R'$ with perfect failure
detectors, using the same type of conversion as 
outlined in Section~\ref{SecFDs}.
We now sketch the conversion. 
Given a run $r \in \R$, we construct a run $f(r)$ such that
\begin{itemize}
\item[P1.] $f(r)_p(0) = (\<\,\>,\ldots, \<\,\>)$;
\item[P2.] if $r_p(m+1) = r_p(m) \cdot e$
and $e$ is not a failure-detector event,
then $f(r)_p(2m+2) =
f(r)_p(2m+1) \cdot e$; 
if $r_p(m+1) = r_p(m) \cdot e$ and $e$ is a failure-detector event or
if $r_p(m+1) = r_p(m)$, then $f(r)_p(2m+2) = f(r)_p(2m+1)$;
\item[P3.] $(f(r))_p(2m+1) = (f(r))_p(2m) \cdot \readpfd{p}{S}$, where $S = \{q: (\R,r,m)
\sat K_p(\quitfla{q})\}$.
\end{itemize}
Thus, in $f(r)$, process $p$'s history is identical to its history in $r$
except that
the failure-detector events in $r$ are deleted in $f(r)$, and,
at each odd step in $f(r)$, $p$'s failure detector reports the
processes
that $p$ knows have crashed at the corresponding point in $\R$.
Now define system $\R^f = \{f(r): r \in \R\}$.
We say that $\R$ {\em can simulate perfect failure detectors\/} if the
\readfdpp\ failure detectors in $\R^f$ are perfect.
We shortly give conditions on $\R$ that guarantee that it can
simulate perfect failure detectors.

As observed by Aguilera, Toueg, and Deianov \citeyear{ATD99}, our definition allows the simulating function
$f$ to be noncomputable.  
Technically, this is not quite right.  The input to $f$ is a run, which
is an infinitary object, so it does not even make sense to 
consider
the computability or noncomputability of $f$.  However, it is easy to
modify $f$ so that its input and output are not complete runs, but
rather prefixes of runs.  Given a prefix of length $m+1$ (i.e., given
$r(0), \ldots, r(m)$ for some run $m$), 
$f$ returns a prefix of length $2m + 2$.
Conditions P1--3 still make sense  with that change.  With that change, it
is clear that $f$ is computable provided that $\{q: (\R,r,m) \sat
K_p(\quitfla{q})\}$ is computable for each $p$, $r$, and $m$.  
While it is possible to 
construct systems where this set is not computable, it will be
computable in any ``reasonable'' system.  That is because whether 
$K_p(\quitfla{q})$ holds at the point $(r,m)$ is typically 
determined by some easily characterizable sequence of events in $p$'s
history.  While it is beyond the 
scope of this paper to characterize when $f$ is computable (it is not
even clear how interesting such a characterization would be), it should
be clear that it typically is computable

The notion of simulation implicitly underlying this 
definition is more general
than, but compatible with, the notion 
of reduction used  by Chandra, Hadzilacos, and Toueg \cite{CHT96}.
For example, in this paper, it is shown
that if consensus can be solved by means of a failure detector (and
there are at most $t < n/2$ failures), then that failure detector can be
transformed to a particular failure detector called
$\Diamond {\cal W}$ (for {\em eventually weak\/}), which satisfies
{\em eventual weak accuracy\/} and 
{\em weak
completeness}; see \cite{CT96} for the precise definition.  Since
consensus can be solved with $\Diamond {\cal W}$ failure detectors,
these failure detectors are viewed as the weakest failure detectors for
consensus.  

The key point is that the results of Chandra, Hadzilacos, and
Toueg do not apply if 
UDC is solved without the use of a failure detector.
However, our notion of simulation does not depend on 
using failure detectors to attain UDC.
Thus, it applies in situations where some other type of oracle is used,
for example, an oracle that gives limited information about which
actions have been initiated, in which case
the reductions of Chandra, Hadzilacos, and Toueg may not apply at all.

We next describe some conditions on a system $\R$ that together will
suffice to show that $\R$ can simulate perfect failure detectors.
Before stating them, we need a definition.
\dfn
A formula $\phi$ local to $q$ is said to be {\em insensitive
to failure by $q$ in $\R$\/} if for all runs $r, r' \in \R$ 
and all times $m, m'$,
if $r'_q(m') = r_q(m) \cdot \quit{q}$, then $(\R,r,m) \sat \phi$
iff $(\R,r',m') \sat \phi$.
\edfn

Now consider the following five conditions on a system $\R$.
\begin{itemize}
\item[A1.] If there exists a run $r_S \in \R$ where all the processes in $S$
crash, and $(r,m)$ is a point in $\R$ such that no process in $\proc -S$
has crashed, then there is a run $r'$ extending $(r,m)$ such that
$F(r') = S$.
\item[A2.] For all runs $r_1, r_2 \in \R$ and times $m$, if $F(r_1) =
F(r_2) =F$ and $(r_1,m) \sim_q (r_2,m)$ for all $q
\notin F$, then there are extensions $r_1'$ and $r_2'$ of 
$(r_1,m)$
and $(r_2,m)$, respectively, such that 
all the processes in $F$ crash by time $m+1$ in $r_1'$ and $r_2'$ and 
$(r_1',m') \sim_q (r_2',m')$
for all $m' \ge m$ and all $q \notin F$.
\item[A3.] The formula $K_q \initfla{p}{\alpha}$ is insensitive to failure by
$q$.
\item[A4.] If $\phi$ is  (a) stable in $\R$, (b) local to some process
$p$ in $\proc$, and (c) insensitive to failure by $p$, then 
for all points $(r,m)$ in $\R$, if there
is some nonempty
$S \subseteq \proc$ 
such that $(\R,r,m) \sat 
\bigwedge_{q \in S} \neg K_q \phi$, then 
there exists a point $(r',m)$ such
that (a) $r'_q(m) = r_q(m)$ for 
all
$q \in S$; 
(b) for all $q \notin S$, there is a (not necessarily strict) prefix $h$
of $r_q(m)$  such that   
either $r'_q(m) = h$ or $r'_q(m) = h \cdot \quit{q}$ and 
$q$ crashes by time $m$ in $r$; and
(c) $(\R,r',m) \sat \neg \phi$.%
\footnote{For those familiar with the notion of distributed knowledge
\cite{FHMV}, note that conditions
(a) and (c) imply that the processes in $S$ do not have  distributed
knowledge of $\phi$.}
\item[A5$_t$.]  For every $S \subseteq \proc$ such that $|S|
\le t$, there exists a run $r_S \in \R$ such that $F(r_S) = S$.
\end{itemize}
We now briefly discuss these conditions and their implications.
A1 essentially says that process failures are independent of other
events.  If it is possible for the processes in $S$ to crash, this may
happen at any time in any run.  A3 says that a process $q$ cannot learn
that $p$ initiated $\alpha$ just by $q$'s crashing.  
A1 and A3 are properties we would
expect to hold of all systems generated
by protocols in the contexts of interest to us.

A2 says that it is possible for all the faulty processes in $r$ that
have not crashed by time $m$ to crash at the next step.
More precisely, if two
points $(r,m)$ and $(r',m)$ are indistinguishable to the correct
processes in $r$, then there are extensions $r_1$ and $r_2$ of these
points that continue 
to be indistinguishable to all the correct processes in $r$, such that
all the faulty processes in $r$ have failed by time $m+1$ in $r_1$ and
$r_2$. 
A2 implicitly assumes that there is no information relevant to the
system beyond what is in the 
correct
processes' states.  
In particular, this means that there cannot be completely reliable
message buffers in the system.
For suppose that $q$
had a message buffer such that once a message was in $q$'s 
buffer, then as long as $q$ did not crash, $q$ would eventually receive
the message.  Consider two runs $r_1$ and $r_2$ such that $(r_1,m)
\sim_q (r_2,m)$,
$F(r_1) = F(r_2)$, and $F(r_1)$ consists of all processes other
than $q$.  Moreover, suppose that there is a message $\m$ in $q$'s
buffer in $(r_1,m)$, but not in $(r_2,m)$.  By 
A2, there are extensions $r_1'$ and $r_2'$ of $(r_1,m)$ and $(r_2,m)$
such that all processes other than $q$ crash in round $m+1$ in both
$r_1'$ and $r_2'$ and $(r_1',m') \sim_q (r_2',m')$ for all $m' \ge m$.
But this is impossible, since $q$ receives $\m$ in $r_1'$ but not in
$r_2'$.
More generally, A2 says that communication is unreliable.  It does not
hold if the network cannot lose all messages that might be in transit at
any given time.

A4 says, among other things, 
that if each of the processes in $S$ considers $\neg \phi$
possible, where $\phi$ is a stable 
failure-insensitive
formula local to some process, then
there is a point where $\neg \phi$ is true that all the processes in $S$
simultaneously consider possible.  
A4 is perhaps the least standard property.
It holds if processes are essentially using a
{\em full-information protocol\/} (FIP)~\cite{Coan,FHMV} and if $\R$ 
places some restrictions on the information they can get from
failure detectors.
With an FIP,
when a process $p$
sends a message to $q$, it sends complete information about its state.
The following example shows that, without FIPs, A4 can fail to be true.
Consider a system $\R$ where processes send messages that are formulas
in the language defined in Section~\ref{SecLanguage}.  Moreover, assume that
every message sent is true at the time that it is sent.  Let $(r,m)$ be
a point in $\R$ such 
that neither $p$ nor $q$ has crashed at $(r,m)$, and at some time
$m'' < m$, 
$q$ sends a message $\m$ to $p'$, which $p'$ receives.  
After receiving $\m$, 
$p'$
sends $p$ a message saying $\quitfla{q} \lor \ess{q}{p'}{\m}$, which $p$
receives by time $m$.  
Further suppose that $p'$ has a perfect failure detector, and there is
another run $r'$ in $\R$ such that $r_p(m) = r'_p(m')$ and, in $r'$,
process $p'$ knows that $q$ has crashed (since its failure detector
reported this) and $q$ does not send $p'$ the message $\m$.
It follows that $(\R,r,m) \sat K_p(\quitfla{q} \lor
\ess{q}{p'}{\m}) \land \neg K_p(\quitfla{q}) \land \neg
K_p(\ess{q}{p'}{\m})$.  Process $p$ knows $\quitfla{q} \lor
\ess{q}{p'}{\m}$ because it received a message from $p'$ saying this,
and messages are known to be truthful in $\R$.  Process $p$ does not
know $\quitfla{q}$ (since $q$ actually has not crashed at the point
$(r,m)$) nor does $p$ know $\ess{q}{p'}{\m}$ (since $\ess{q}{p'}{\m}$ is
not true at $(r',m')$). 
But then A4 does not hold in $\R$ for $\phi = \ess{q}{p'}{\m}$ and
$S = \{p\}$.  For suppose it did hold.
Then there must be a point $(r'',m)$ in $\R$ such that
(a) $r''_p(m) = r_p(m)$, (b) $r''_q(m)$ is a prefix of $r_q(m)$ 
(since $q$ does not crash in $r$), 
and (c)
$(\R,r'',m) \sat \neg \ess{q}{p'}{\m}$.  Since 
$(\R,r,m) \sat K_p(\quitfla{q} \lor
\ess{q}{p'}{\m})$, it follows that $(\R,r'',m) \sat \quitfla{q}$,
violating the assumption that $r''_q(m)$ is a prefix of $r_q(m)$.

In this example, 
$p'$ did not tell $p$
all it knew, which is precisely what cannot happen with a
full-information protocol.
Assuming that $\R$ is generated by an FIP, under reasonable assumptions
about the runs in $\R$ (discussed below), $\R$ will satisfy A4.  To see
why, observe that given $(r,m)$ 
and $\phi$ as in the hypotheses of A4,  
we can construct the run $r'$ as follows.
First note that 
$(\R,r,0) \sat \neg\phi$, for otherwise, since
$\phi$ is stable and local to $p$,
$\phi$ would be true at all points in $\R$ and so would $K_q \phi$
for all $q \in \proc$.
Thus, let
$m_p$ be the first 
time in $r$ where $\phi$ becomes true. 
If $m_p > m$, then take $r' = r$ and $S = \proc$;  
A4 trivially holds in this case. 
If $m_p \le m$, let
$S\subseteq \proc$
be the processes that do not know $\phi$ at $(r,m)$.  If processes are
following a full-information protocol, there can be no chain of messages
from $p$ to a process $q \in S$ between times $m_p$ and $m$ in $r$, for if
there were, $q$ would know $\phi$ at $(r,m)$.%
\footnote{There is a message chain from $p$ to $q$ between $m_p$ and $m
> m_p$ if there 
is a sequence of messages $\m_1, \ldots, \m_k$ and processes $p_1,
\ldots, p_{k+1}$ such that 
(a) $\m_i$ is sent by $p_i$ to $p_{i+1}$
and is received,
(b) $p_{i+1}$ sends $\m_{i+1}$ after receiving $\m_i$, 
(c) $p = p_1$, (d) $q = p_{k+1}$, (e) $p$ sends
$\m_1$ at or after $m_p$, and (f) $q$ receives $\m_{k+1}$ at or before
$m$.  If the processes follow a full-information protocol, then
when $p_{i+1}$ receives $\m_i$, $p_{i+1}$ all the stable facts that $p_i$
knew when $p_i$ sent $\m_i$.} 
{F}or each process $q \in \proc$, let
$m_{q}$ be the least time at or
before $m$ at which
there is a message chain from $p$ to $q$ in $r$ between $m_p$ and $m_q$,
if there is such a time;
otherwise, we take 
$m_{q} = m+1$.
Note that for $q \in S$, we have $m_q = m+1$.
We then construct $r'$ so that $r'_q(m') = r_q(m')$ for 
$m' \le m_q - 1$;
if $q$ does not crash in $r$ between times $m_q$ and $m$ inclusive, then
$r_q'(m') = r_q(m_q-1)$ for $m' \ge m_q$; otherwise,   
$r_q'(m') = r_q(m_q-1)\cdot \quit{q}$ for $m' \ge m_q$.  
By construction, we have $r_q(m) = r'_q(m)$ for $q \in S$.  For $q' \notin
S$, we have that $r_{q'}'(m)$ is either $r_{q'}(m_q-1)$ or $r_{q'}(m_q-1) \cdot
\quit{q'}$.  The reason
we need to add $\quit{q'}$ is that the failure detector of
some process $q \in S$ might report that $q'$ fails in $r$.  Since
$r_q(m) = r'_q(m)$, if $q$'s failure detector is accurate, it must be
the case that $q'$ also fails in $r'$.  
As long as $r' \in \R$, it is easy to see that the point $(r',m)$
satisfies the requirements of (this instance of) A4.  For by
construction, $r_q(m) = r'_q(m)$ for $q \in S$; and for $q' \notin S$,
the construction guarantees that either $r'_{q'}(m) = r_{q'}(m)$ or
$r'_{q'}(m) = r_{q'}(m_q)\cdot \quit{q'}$.   By choice of $m_p$, we
have that $(\R,r,m_p - 1) \sat \neg \phi$.
Note that if $(\R,r,m) \sat \neg \quitfla{p}$ then
$r'_p(m) = r_p(m_p - 1)$; otherwise, either $r'_p(m) = r_p(m_p-1)$ or
$r'_p(m) = r_p(m_p-1) \cdot \quit{p}$. Since
$\phi$ is insensitive to failure by $p$, in either case,
we have that $(\R,r',m) \sat \neg \phi$.  
Thus, $(r',m)$ satisfies the requirements of A4.

This argument shows is that as long as it is the case that, for each
formula $\phi$ and 
point $(r,m)$ satisfying the hypotheses of A4, there is a run $r' \in \R$ as
constructed above (actually, it suffices that there is a run in $\R$ that
extends $(r',m)$), then $\R$ satisfies A4.  Thus, for example, it cannot
be the case that the failure detector reports in $\R$ are correlated
with message delivery, so that a report from
a failure detector saying that $p$ is faulty is accurate iff $p$ did not
receive a message from $q$.  We do not attempt to completely
characterize the conditions under which $\R$ satisfies A4 here.

A5$_t$ says that any subset of processes of size at most $t$ may fail in
some run.  
This is a standard assumption in the literature.
Note that A5$_t$ implies A5$_{t'}$ if $t \ge t'$.

Theorem~\ref{UDCchar} below shows
that if $\R$ 
attains UDC and satisfies 
A1--A4 and
A5$_n$ (or A5$_{n-1}$) 
and one other quite innocuous condition, 
then 
$\R$ can simulate perfect failure detectors.
The ``innocuous condition'' is the following: Clearly any solution to UDC
should allow a process to initiate any of its actions at any time.  
To guarantee that $\R$ can simulate perfect failure detectors, 
it is
necessary that, in each run of $r$, 
the correct processes (if there are any) initiate actions
infinitely often.  
That is, for all runs $r$, if $F(r) \ne \proc$, then for all
times $m$, some correct process in $r$ initiates an action
after $(r,m)$.
Intuitively, if actions are initiated infinitely often, the correct
processes will need to be able to detect
failures in order to attain UDC repeatedly.  On the other hand, if the
correct processes do
not initiate actions after some point, there will be no need for them to
detect failures after this point.
To see the need for this
condition, suppose that no actions are initiated after time 17, even
though there are some correct processes in $\R$ and UDC is attained for
all these actions by time 25.  Now consider a process $q$ that fails
after time 17.  There is no need for processes to know that $q$ has
failed, since UDC is not required after time 25.

To summarize, 
our result can be viewed as saying that under some 
relatively innocuous assumptions (A1--A3 and the assumption that correct
processes initiate actions infinitely often),
if any subset of processes may fail (A5$_n$) and the processes
are telling each other as much as they can (A4), then 
being able to attain UDC is
tantamount to being able to simulate 
perfect 
failure detectors.
Before proving Theorem~\ref{UDCchar},
we prove two preliminary results.  The first shows that, in the contexts
of interest to us, weak accuracy  and strong accuracy are equivalent.
The second provides a characterization of the facts that must
be known by a process before it can perform 
a coordination action 
$\alpha$.  Specifically, a
process must know that if there are any correct processes at all, then one
of these knows that $\alpha$ has been initiated.

\pro\label{strong=perfect}  If $\R$ satisfies A1 and A5$_{n-1}$ 
then $\R$ satisfies weak accuracy iff $\R$ satisfies strong accuracy.
\epro
It 
follows from Proposition~\ref{strong=perfect} that if
$\R$ satisfies A1 and A5$_{n-1}$, then $\R$ has strong failure detectors
iff $\R$ has perfect failure detectors.  
(Since A5$_n$ implies A5$_{n-1}$, this is {\em a fortiori\/} the case if
$\R$ satisfies A1 and A5$_n$.)

\gobble{
The next proposition is key to the analysis.
If UDC holds, when a correct process $p$
initiates $\alpha$, then $p$ must eventually perform $\alpha$.  At this
point, $p$ must know that eventually every correct process $q$ will also
perform $\alpha$ (and thus must eventually know that $p$ initiated
$\alpha$).  As we now show, this can happen only if $p$ knows that
if there are any non-faulty processes at all, then
there is some process that is correct throughout the run that knows that
$p$ initiated $\alpha$.
}

\pro\label{PropAccount}
If $\R$ satisfies A1, A2, and A4, then 
$$\begin{array}{l}
\R\sat
\bigwedge_{p, p'\in\proc}\bigwedge_{\alpha\in \A_{p'}}
\biggl[ 
K_p\Bigl(
\initfla{p'}{\alpha} \land
\bigwedge_{q \in \proc} \Diamond( K_q
\initfla{p'}{\alpha} \vee \quitfla{q}) \Bigr)  \\ 
\mbox{\ \ \ \ \ \ \ \ \ }\Rightarrow 
K_p \biggl(\bigvee_{q\in \proc} \Box \neg \quitfla{q} \rimp 
\bigvee_{q\in \proc} \Bigl(K_q
\initfla{p'}{\alpha} \land \Box \neg \quitfla{q}\Bigr)\biggr)
\biggr].
\end{array}\]
\epro

We 
are now ready to state our theorem.

\thm\label{UDCchar} Suppose $\R$ is the system generated by a protocol
that attains UDC, $\R$ satisfies 
A1--A4 and
A5$_{n-1}$,
and for each run $r \in \R$,
if $F(r) \ne \proc$, then infinitely many actions are initiated in $r$ 
(i.e., infinitely many events of the form $\init{p}{\alpha}$ appear in $r$).
Then the system $\R^f$ has perfect failure detectors. \ethm

There are two issues worth noting regarding Theorem~\ref{UDCchar}.%
\footnote{We thank one of the reviewers of this paper for pointing out
these issues and encouraging us to discuss them.}
First, the alert reader may have noticed an apprarent contradiction in our
results.  Proposition~\ref{UDCok} states the UDC can be attained in
contexts where communication is reliable, without using failure detectors.
Theorem~\ref{UDCchar} states that, if UDC can be attained, then perfect
failure detectors can be simulated.  Thus, perfect failure detectors can
be simulated in systems where 
communication is reliable.  Since it is
well known that Consensus can be attained with perfect failure detectors,
this suggests that Consensus can be attained in systems where
communication is reliable, regardless of the number of process
failures.  But this is well known to be false \cite{FLP}.  

Our results are correct.  We escape from the contradiction because, as
we noted earlier, A2 specifically precludes reliable communication.
Thus, Theorem~\ref{UDCchar} does not apply to systems of the type
considered in Proposition~\ref{UDCok}.  This observation does emphasize
that our main results apply only to systems where communiation is
unreliable.  

Second, the assumption that infinitely many actions must be initiated in
each run of $\R$ in Theorem~\ref{UDCchar} may strike some readers as
unduly strong
(although it can be argued that a service should expect to operate
indefinitely and therefore handle infinitely many requests).  In any case,
the theorem can be rephrased in a way that
might make it more palatable.  Consider a context that allows solutions
to UDC and satisfies A1--A4 and A5$_{n-1}$.  Then there is a joint
protocol $(P_1,\ldots, P_n)$ that, when run in that context, generates a
system $\R$ such that $\R^f$ has perfect failure detectors.  The proof
of this result is essentially identical to that of
Theorem~\ref{UDCchar}: the joint protocol $(P_1, \ldots, P_n)$ is one
where each 
process that does not crash initiates an infinite number of actions.
Indeed, every joint protocol where each process that does not crash
initiates an infinite number of actions generates a
system $\R$ where $\R^f$ has perfect failures detectors.  Thus, the
result really shows that in contexts where UDC can be solved, UDC can be
used to generate perfect failure detectors.

\section{Generalized Failure Detectors}\label{SecGen}

Theorem~\ref{UDCchar} shows that if 
at as many as  $n-1$ processes can fail, then
UDC essentially requires perfect 
failure detectors.  On the other hand, as Gopal and Toueg
\citeyear{GT-WDAG89} show,
UDC is attainable without using failure detectors in contexts where
there are less than  $n/2$ failures.
We now generalize both of these results,
characterizing the type of failure detector needed to attain UDC if there
is a bound of $t$ on the number of possible failures, for all values of
$t$.

A {\em generalized failure detector\/} reports that (it suspects that) at
least $k$ processes in a set $S$ are faulty.%
\footnote{Despite the name, generalized failure detectors are a special
case of failure detector as defined in Section~\ref{SecFDs}, as well as
being a special case of the failure detectors defined by Aguilera,
Toueg, and Deianov \citeyear{ATD99}.}
As discussed in the Introduction, such generalized failure detectors are
appropriate when processes can observe faulty behavior in some
component(s) without being able to tell which processes in the component
are actually faulty.  We model such generalized suspicions by using
events
of the form $\readfd{p}{S,k}$, with $k \le |S|$.%
\footnote{Again, it is not necessary that the report of the failure
detector has the form $(S,k)$.  We can define $g$-generalized failure
detector whose reports can be mapped to pairs $(S,k)$.  For ease of
exposition, we do not bother doing this.}
We are interested in generalized failure
detectors that give useful information.  Of course, what is ``useful'' may
depend on the application.  Given a system $\R$ and an upper bound of $t$
on the number of failures that may occur in a run $r$ of $\R$, we say that
$\readfd{p}{S,k}$ is a {\em $t$-useful failure-detector event for
$r$\/} if (a) $F(r) \subseteq S$, 
(b)
$n - |S| > \min(t,n-1) -k$ (or, equivalently, $k > |S| - n +
\min(t,n-1)$), 
and (c) $k \le |S|$.
Intuitively, if a generalized failure detector is ``good'', then some of
its reports are $t$-useful failure events.
Note that if $p$ learns at the point $(r,m)$ that there are $k$ faulty
processes in $S$ and $n - |S| > \min(t,n-1)-k$, then
$p$ can conclude that, if there are any correct processes at all in
$r$, then one of the processes in $\proc - S$ is
correct at $(r,m)$ (although it may not know which one).
Just knowing that some processes in a set are correct is not 
useful in general.  For example, if $t < n$, then all processes know
that at least $n 
- t$ processes in $\proc$ are correct.  As we shall see, what makes this
fact useful is that $F(r)\subseteq S$.
A generalized failure detector in $\R$ is {\em $t$-useful\/} if for all
$r \in \R$ and processes $p$, we have: 
the following:

\begin{description}
\item[{\em Generalized Strong Accuracy:}] if $\readfd{p}{S,k}$ is in
$r_p(m)$, then there is a subset $S' \subseteq S$ such that $|S'| = k$ and
for all $q \in S'$, we have that $\quit{q}$ is in $r_q(m)$.

\item[{\em Generalized Impermanent Strong Completeness:}] if $p$ is
correct, then there 
is
a $t$-useful failure-detector event for $r$ in
$r_p(m)$, for some $m$.

\end{description}

Note that it is trivial to construct a $t$-useful failure detector in a
context with at most $t$ failures if $t < n/2$: 
for each $S \subseteq \proc$ with $|S| = t$, output $(S,0)$ infinitely often.
Suspecting no processes in any
subset $S$ trivially satisfies generalized strong accuracy, and in every
run $r$ at least one $t$-sized subset of \proc\ must contain $F(r)$.
Whenever $F(r) \subseteq S$, then $(S,0)$ is a $t$-useful
failure-detector event.

Also note that if $\readfd{p}{S,k}$ is an $(n-1)$-useful or $n$-useful
failure-detector event,
then we must have $|S| = k$, since the only way to have $k > |S|-1$
is to have $k = |S|$.
Thus, we can easily convert an $n$-useful or
$(n-1)$-useful generalized failure detector to a perfect failure detector,
by just reporting 
$\readfd{p}{S'}$ at time $m$ in run $r$ if $S'$ is the union of the sets
$S$ such that the
generalized failure detector has reported $\readfd{p}{S,k}$ with $|S|=k$
prior to time $m$.  
Conversely, we can easily convert a perfect failure detector to an
$n$-useful (and hence $(n-1)$-useful)
failure detector.  Given a history for process $p$, we
simply replace each event $\readfd{p}{S}$ by the event $\readfd{p}{S',k}$,
where $S'$ is the union of $S$ together with all the sets that appeared in
failure-detector events of the perfect failure detector earlier in the
history, and $k = |S'|$.  It is easy 
to see that this gives an $n$-useful failure detector.  Thus, the following
result generalizes Proposition~\ref{UDC} and Gopal and
Toueg's result.

\pro\label{UDC1} There is a protocol that attains UDC in a context
with a bound of $t$ on the number of failures and with $t$-useful
generalized failure detectors.
\epro

Since, as observed earlier, it is trivial to construct a $t$-useful
failure detector in a context with at most $t$ failures, if $t < n/2$,
we get the following result of Gopal and Toueg \citeyear{GT-WDAG89} as
an immediate corollary to Proposition~\ref{UDC1}.
\cor\label{cor:UDC1} If $t < n/2$, then there is a protocol that attains
UDC without failure detectors.
\ecor

We want a converse to Proposition~\ref{UDC1} 
that generalizes
Theorem~\ref{UDCchar}.   We show that if processes can perform UDC
in a context with a bound $t$ on the number of failures, then
$t$-useful generalized failure detectors can be simulated in that context.

Given system $\R$, construct system $\R^{f'}$ as follows.  
Fix an order $S_0, \ldots, S_{2^n-1}$ of the subsets of $\proc$.
Let $\R^{f'} = \{f'(r): r \in \R\}$ where, for each run $r \in \R$, $f'(r)$ is
constructed  
exactly as $f(r)$ in Section~\ref{SecKA}, except that P3 is replaced by the
following condition.
\begin{itemize}
\item[P3$'$.] $(f'(r))_p(2m+1) = (f'(r))_p(2m) \cdot \readpfd{p}{S_l,k}$, 
where $l$ is the length of the
history $r_p(m+1)$ mod $2^n$ and 
$$\mbox{$k = \max \{ k' : (\R,r,m) \sat K_p( k'$ processes in $S_l$ have
crashed$)\}$.}$$ 
\end{itemize}

\thm\label{UDCchar1} Suppose $\R$ is the system generated by a protocol
that attains UDC in a context with at most $t$ failures, $\R$ satisfies
A1--A4 and A5$_t$, and for each run $r \in \R$, 
if $F(r) \ne \proc$, then infinitely many actions are initiated in $r$. 
Then $\R^{f'}$ has $t$-useful
generalized
failure detectors. \ethm

As with Theorem~\ref{UDCchar}, we can restate Theorem~\ref{UDCchar1} to
say that in any context with at most $t$ failures where UDC can be
attained, there is a joint protocol $\vec{P}$ that generates a system
$\R$ such that $\R^{f'}$ has $t$-useful failure detectors.

\section{Conclusions}
\label{sec:conclusions}

\begin{table*}
\begin{center}
\begin{tabular}{|ll||l|l|l|} \hline
                  & & $0< t < n/2$ & $n/2 \leq t < n-1$ & $n-1 \leq t \leq
n$ \\ \hline
Reliable channels &UDC &no FD & no FD & no FD \\
                  &consensus       &$\Diamond{\cal W}$~~$\dagger$
&Strong &Perfect~~$\dagger$ \\ \hline
Unreliable channels &UDC &no FD   &$t$-useful~~$\dagger$
&Perfect~~$\dagger$ \\
& consensus      &$\Diamond{\cal W}$~~$\dagger$  &Strong
&Perfect~~$\dagger$ \\ \hline
\end{tabular}
\caption{The type of failure detector needed for UDC vs.~consensus;
$\dagger$ indicates optimality.}
\end{center}
\end{table*}

We have shown that the problem of Uniform Distributed Coordination in
asynchronous systems varies in its complexity both with communication
guarantees and
with
the number of failures that must be tolerated (see
Table~1).
Unlike consensus
(or nUDC, for that matter),
UDC is sensitive to communication guarantees
in the contexts that we consider in this paper.
This is significant since UDC is likely the only acceptable reliability
guarantee for many wide-area 
and collaborative mobile 
applications, precisely where reliable
communication cannot be assumed.

Note that we have completely characterized the type of failure detector
required to attain UDC for all values of $t$.  For consensus, it is known
that $\Diamond {\cal W}$ is necessary and sufficient if $t < n/2$.
(Recall that in this case no failure detectors at all are necessary to
attain UDC.)  While strong (actually,
impermanent-strong) failure detectors suffice for consensus for $n/2 \le
t < n$, there is no characterization of exactly the type of failure
detector that is required.  The notion of $t$-useful failure detectors
defined here may prove useful in that regard.  We leave exploring this
issue for future work.

As we mentioned in the introduction, in a paper written in response to
the conference version of  this paper, Aguilera, Toueg, and Deianov
\citeyear{ATD99} provided an elegant alternative characterization of the
weakest failure detector required for UDC.%
\footnote{Actually, their results are
given for URB, {\em uniform reliable broadcast}, but URB and UDC are
isomorphic problems; the {\em init\/} and {\em do\/} in UDC correspond
to {\em broadcast\/} and {\em deliver\/} in URB.}  They show that the
weakest failure detector for this problem is one that satisfies strong
completeness and a notion of accuracy even weaker than what we have
called weak accuracy: if there is a correct process, then at all times,
some correct process is not suspected (but a different correct process
may be the one that is not suspected at every time).  They show that if
UDC can be solved with a failure detector $F$, then $F$ can be reduced
(i.e., effectively transformed) into this weakest failure detector.  
Technically, this result is incomparable to our result.  On the one
hand, it is stronger, in that it gives a failure detector that solves
UDC in all contexts (not just ones satisfying A1--A4 and A5$_t$, which
are the only ones considered in this paper), and is in a precise sense
the weakest failure detector needed to solve UDC.  On the other hand, as
we observed in the discussion preceding
Theorem~\ref{UDCchar}, because our results do not proceed by reduction of
one failure detector to another,  our results apply even in cases where some
technology other than failure detectors is used to solve UDC.

\gobble{
At first blush, unreliable channels seem to cause severe hardship for UDC,
but not for consensus.  Weak accuracy ensures the existence of a correct
process, say $c$, that is never suspected.  Since $c$ is correct, the
difference between unreliable and reliable channels is irrelevant: under
our fairness assumption, each distinct message from $c$ will reach other
correct processes after some unknown but finite number of retransmissions.
Unreliable channels may lengthen the actual transmission delay of a given
message but do not impose further requirements on weak accurate failure
detectors, which are already clever enough never to suspect some correct
process despite asynchronous transmission delays.  On the other hand,
suspicions
arise in consensus solutions only as an alternative to receiving
(or not) a message.  We can model the loss of an expected message from $p$
to $q$ with $q$'s failure detector suspecting $p$; for processes other than
$c$, erroneous suspicions are permitted, so these message losses can be
tolerated.}

\section*{Appendix -- Proofs of Propositions and Theorems}
\opro{nUDC} There is a protocol that attains nUDC without the use of failure
detectors in every context where communication is fair (although possibly
unreliable), even if there is no bound on the number of failures.

\eopro

\medskip

\prf We just sketch the protocol here, since it is so simple.  Whenever a
process $p$ wants to attain nUDC of action $\alpha$ (i.e., if
$\init{p}{\alpha}$ is in $p$'s history) $p$ goes into a special {\sf
nUDC($\alpha$)} state.  If a process is in an {\sf nUDC($\alpha$)} state,
it performs $\alpha$ and sends an 
{\em $\alpha$-message\/} repeatedly to all other
processes (which, intuitively, tells them to perform $\alpha$).  If a
process receives an $\alpha$-message, it
goes into an {\sf nUDC($\alpha$)} state, if it has not already done so.  It
is easy to see that this protocol attains nUDC.\footnote{This protocol,
like most of the others we present in this paper, does not have any
mechanism for termination.  Processes keep sending messages forever.  Since
message communication is unreliable, it is not hard to show that there is
in fact no protocol that attains nUDC and terminates.  We can deal with
this problem by adding a {\em heartbeat\/} mechanism \cite{ACT97}, but this
issue is beyond the scope of this paper.}  \eprf

\opro{UDCok}  There is a protocol that attains UDC without the use of
failure detectors in every context where communication is reliable, even
if there is no bound on the number of failures. 
\eopro
\medskip
\prf We proceed just as in the proof of Proposition~\ref{nUDC}, except that
before performing the action $\alpha$, a process simply sends a message to
all other processes telling them to perform $\alpha$ and inform all other
processes if they have not already done so.  
More precisely, if $\init{p}{\alpha}$ is in $p$'s history, $p$ goes into
a special {\sf UDC($\alpha$)} state.  If a process is in a 
{\sf UDC($\alpha$)} state, it sends an $\alpha$-message to all processes
and then performs $\alpha$.  If a process receives an $\alpha$ message,
it goes into a UDC-state if it has not already done so.
Since a process $q$ performs $\alpha$
only after sending out an $\alpha$-message to all 
processes and, by assumption, communication is reliable, if $q$ performs
$\alpha$, then other correct processes will receive the message, and 
thus  
also perform $\alpha$, even if $q$ crashes.  \eprf

\opro{UDC} There is a protocol that attains UDC in 
every context with
strong failure detectors,
even if there is no bound on the number of failures.
\eopro
\medskip
\prf The proof is similar in spirit to that of Proposition~\ref{nUDC}.
Whenever a process wants to attain UDC of action $\alpha$, it goes into a
special {\sf UDC($\alpha$)} state.  If a process $p$ is in a {\sf
UDC($\alpha$)} state, it sends an $\alpha$-message repeatedly to all other
processes (telling them to perform $\alpha$).
Process $p$ performs $\alpha$ if
it is in a {\sf UDC($\alpha$)} state and if,
for every process $q$, $p$
receives an acknowledgment from $q$ to its $\alpha$-message or $p$'s
failure detector says 
or has said
that $q$ is faulty.  However, $p$ continues to send
$\alpha$-messages (even after performing $\alpha$) to all processes from
which it has not received an acknowledgment
until it has received an
acknowledgment from {\em all\/} processes (which may never happen).%
\footnote{If $p$ has a strongly accurate failure detector rather than just
a weakly accurate failure detector, it can actually stop sending messages
after performing $\alpha$.  This follows from the proof of
Proposition~\ref{UDC}.}  Every time a process $q$ receives an
$\alpha$-message from $p$, $q$ sends an acknowledgment to $p$; it also goes
into a {\sf UDC($\alpha$)} state if it has not already done so.

To show that this protocol attains UDC, it suffices to show that, in every
run, (1) if a process $p$ is in a {\sf UDC($\alpha$)} state, then $p$ will
eventually perform $\alpha$ or crash and (2) if $p$ performs $\alpha$ then
every correct process performs $\alpha$.  To see that (1) holds,
suppose that $p$ is in a {\sf UDC($\alpha$)} state in run $r$
and does not crash.
Suppose, by way of contradiction, that $p$ does not perform $\alpha$ in
run $r$.
That means that there must be some process $q$ such that $p$'s failure
detector never reports $q$ as faulty and $p$ does not receive an
acknowledgment from $q$.  Since $p$ has a strong failure
detector, if $q$ is faulty, then at some point in $r$, $p$ must receive
a report to 
this effect from its failure detector.  Thus, $q$ must be correct in $r$.
Since $p$ repeatedly sends an $\alpha$-message to $q$, 
by R5, $q$ must receive the message infinitely often.  That means it
sends an acknowledgment back to $p$ infinitely often.  By R5 again, $p$
must receive the acknowledgment, contradicting the assumption that it
does not.  

To see that (2) holds, 
first note that it holds vacuously if there are no processes correct in $r$.
If there is some process that is correct in $r$, 
then
since $p$ has a weakly accurate
failure detector, there is some correct process, say $q^*$, that $p$ never
suspects.  Thus, if $p$ performs $\alpha$, it must receive an
acknowledgment from $q^*$ to its $\alpha$-message.  Hence, 
$q^*$ goes into a
{\sf UDC($\alpha$)} state and never crashes, so (1) implies that it also
performs 
$\alpha$.  
Since $q^*$ is correct, all correct processes 
eventually receive an $\alpha$-message from $q^*$ and
so 
perform
$\alpha$.  \eprf

\opro{strong=perfect}  If $\R$ satisfies A1 and A5$_{n-1}$, 
then $\R$ satisfies weak accuracy iff $\R$ satisfies strong accuracy.
\eopro
\medskip

\prf
Let $\R$ satisfy A1, A5$_{n-1}$, and weak accuracy.  If $\R$
does not satisfy strong accuracy, then there is a point $(r,m)$ 
and processes $p$, $q$ such that $q \in \psuspects{p}(r,m)$ and $q$ has
not failed in $r$. Let $S' = \proc - 
\{q\}$.  By A5$_{n-1}$, there is a run $r'$
where all the processes in $S'$ fail.  Thus, by A1, there is a run $r''$
extending $(r,m)$ such that all the processes in $S'$ fail in $r''$.  
It follows that
$q$ is the only correct process in $r''$.  By weak accuracy, we must have
that $q$ is never suspected as faulty in $r''$, contradicting the
assumption that it is in fact suspected by $p$.  \eprf

\opro{PropAccount}
If $\R$ satisfies A1, A2, and A4, then 
$$\begin{array}{l}
\R\sat
\bigwedge_{p, p'\in\proc}\bigwedge_{\alpha\in \A_{p'}}
\biggl[ 
K_p\Bigl(\initfla{p'}{\alpha} \land
\bigwedge_{q \in \proc} \Diamond( K_q
\initfla{p'}{\alpha} \vee \quitfla{q}) \Bigr)  \\ 
\mbox{\ \ \ \ \ \ \ \ \ }\Rightarrow 
K_p \biggl(\bigvee_{q\in \proc} \Box \neg \quitfla{q} \rimp 
\bigvee_{q\in \proc} \Bigl(K_q
\initfla{p'}{\alpha} \land \Box \neg \quitfla{q}\Bigr)\biggr)
\biggr].
\end{array}\]
\eopro

\prf Suppose, by way of contradiction, that 
for some $p, p' \in \proc$ and $\alpha \in \A_{p'}$, we have that 
\begin{equation}\label{eq:contra1} \begin{array}{ll}
(\R,r,m) \sat 
\ K_p
\Bigl( \initfla{p'}{\alpha} \land \bigwedge_{q \in \proc} \Diamond( K_q
\initfla{p'}{\alpha} \lor \quitfla{q})\Bigr)~~ \land \\ 
\mbox{\ \ \ \ \ \ \ \ \ \ \ \ \ \ \ }\ \ \neg K_p
\biggl(\bigvee_{q\in \proc} \Box \neg \quitfla{q}
\rimp \bigvee_{q \in
\proc} \Bigl(K_q \initfla{p'}{\alpha} \land \Box \neg \quitfla{q}\Bigr)
\biggr).  \end{array} \end{equation}

\noindent Then there must be a point $(r^1,m') \sim_p (r,m)$ such that
$$(\R,r^1,m') \models \initfla{p'}{\alpha} \land
\bigvee_{q\in \proc} \Box \neg \quitfla{q} \land 
\bigwedge_{q \in \proc}
\Bigl(\Box \neg \quitfla{q} \rimp \neg K_q \initfla{p'}{\alpha}\Bigr).
$$

\noindent
We have $(\R,r^1,m') \sat \bigwedge_{q \notin F(r^1)} \neg K_q
\initfla{p'}{\alpha}$.  Since $p'$ knows that it initiated $\alpha$
at $(r^1,m')$, we must have $p' \in F(r^1)$.  Moreover, $F(r^1) \ne \proc$, 
because $(\R,r^1,m') \sat \bigvee_{q\in \proc} \Box \neg \quitfla{q}$.

Let $S - \proc - F(r^1)$.
By A4 with $\phi =_{\rm def} \initfla{p'}{\alpha}$, 
there exists a point $(r^2,m')$ such that $(r^2,m') \sim_{q}
(r^1,m')$ for $q \in S$ and $(\R,r^2,m') \sat \neg
\initfla{p'}{\alpha}$.  
For all $q \in S$), we have that $(r^2,m') \sim_q
(r^1,m')$.  It follows that no process in $S$ has crashed by
$(r^2,m')$. 
By A1, there exists a run $r^3$ extending $(r^2,m')$
such that $F(r^3) = F(r^1)$.  Since $r^3$ extends $(r^2,m')$, we must have
$r^1_q(m') = r^3_q(m')$ for all $q \in S$.  By A2, there exist
runs $r^4$ and $r^5$ extending $r^1$ and $r^3$, respectively, such that
$r^4_q(m'') = r^5_q(m'')$ for $m'' \ge m'$. Moreover, all the processes in
$F(r^1)$ (and, in particular, $p'$) crash by time $m' + 1$ in $r^4$ and
$r^5$.  Thus, the event $\initfla{p'}{\alpha}$  does not appear in $r^5$,
which means that $(\R, r^5,m') \sat \bigwedge_{q \in S}
\never K_q
\initfla{p'}{\alpha}$.  Since $r^4$ and $r^5$ are indistinguishable to such
$q$ from $m'$ onward, we have $(\R, r^4, m') \sat \bigwedge_{q \in S}
\never K_q \initfla{p'}{\alpha}$.
Since $(r,m) \sim_p (r^1,m')$ and $r^4$ extends $(r^1,m')$, we must have
$(r,m) \sim_p (r^4,m')$.  Hence, we have $(\R,r,m) \sat \neg K_p
\Bigl(\Diamond( K_{q} \initfla{p'}{\alpha} \vee \quitfla{q})\Bigr)$ for $q
\in S$.  This gives the desired contradiction to
(\ref{eq:contra1}).  \eprf 

\othm{UDCchar} Suppose $\R$ is the system generated by a protocol that
attains UDC, $\R$ satisfies A1--A4 and A5$_{n-1}$, and for
each run $r \in \R$, 
if $F(r) \ne \proc$, then infinitely many actions are initiated in $r$ 
(i.e., infinitely many events of the form $\init{p}{\alpha}$ appear in $r$).
Then the system $\R^f$ has perfect failure detectors. 
\eothm
\medskip

\prf It is immediate from the construction that $p$ crashes in $(r,m)$ iff
$p$ crashes in $(f(r),2m)$.  It easily follows that $p$'s failure detector
satisfies strong accuracy.  To show that it satisfies strong completeness,
suppose that $p$ is correct and $q$ fails in run 
$f(r) \in \R^f$ and hence
also in run $r \in \R$.  Since infinitely many actions are initiated in
$r$ (and hence $f(r)$), there must be some action $\alpha$ initiated by
some correct process, say $p'$, in $f(r)$ after $q$ has failed.  
Since $\R$ satisfies UDC, by DC1 and DC2, $p$ must eventually
perform $\alpha$ in run $r$, say at time $m$.  Moreover, by DC2, $p$ knows
that, for each process $q'$ (and, in particular, $q$), $q'$ must eventually
either crash or must perform $\alpha$.  Using DC3, it easily follows that
we must have $$(\R,r,m) \sat K_p\Bigl(\initfla{p'}{\alpha}
 \land  \bigwedge_{q' \in 
\proc} \Diamond (K_{q'} \initfla{p'}{\alpha} \vee \quitfla{q'})\Bigr).$$  
Since
$\R$ satisfies A1, A2, and A4 by assumption, it follows from
Proposition~\ref{PropAccount} that
\begin{equation}\label{eq:contra2}
(\R,r,m) \sat 
K_p \biggl(\bigvee_{q'\in \proc} \Box \neg \quitfla{q'} \rimp 
\bigvee_{q' \in \proc} \Bigl(K_{q'}
\initfla{p'}{\alpha} \land \Box \neg \quitfla{q'}\Bigr)\biggr).
\end{equation}

By way of contradiction, suppose
that $(\R,r,m)\sat \never K_p
\quitfla{q}$.  Since $q$ crashes in $r$ before $p'$ initiates $\alpha$, 
it is easy to show that $(\R,r,m) \sat \neg K_q \initfla{p'}{\alpha}$.
Thus, 
there must 
exist a point $(r^1,m') \sim_p (r,m)$ such that $(\R,r^1,m') \sat \neg
\quitfla{q} \land \neg K_p K_q \initfla{p'}{\alpha}$.  Since
$K_q\initfla{p'}{\alpha}$ is stable, local to $q$,
and (by A3) insensitive to failures by $q$, 
by A4, there must
exist a point $(r^2,m') \sim_p (r^1,m')$ such that $r^2_q(m')$ is a prefix
of $r^1_q(m')$ and $(\R,r^2,m') \sat \neg K_q \initfla{p'}{\alpha}$.  
Since $r^2_q(m')$ is a prefix of $r^1_q(m')$, it is easy to show
that $(\R,r^2,m') \sat \neg
\quitfla{q}$.  Thus, $(r^2,m') \sim_p
(r,m)$ and $(\R,r^2,m') \sat
\neg\quitfla{q} \land \neg K_q\initfla{p'}{\alpha}$. 

By A5$_{n-1}$ and A1, there is a run $r^3$
extending $(r^2,m')$ such that all processes except $q$ fail
in $r^3$.
Since $(r^3,m') \sim_p (r,m)$, and
$$(\R,r^3,m') \sat \never \quitfla{q} \land \neg K_q \initfla{p'}{\alpha}
\land  \bigwedge_{q' \in \proc-\{q\}} \Diamond \quitfla{q'} .$$ 
Since $(r,m) \sim_p (r^2,m') \sim_p (r^3,m')$, this
contradicts~(\ref{eq:contra2}).  \eprf

\opro{UDC1} There is a protocol that attains UDC in a context with a bound
of $t$ on the number of failures and with $t$-useful generalized failure
detectors.  \eopro
\medskip

\prf To attain UDC of action $\alpha$, a process goes
into a special {\sf UDC($\alpha$)} state.  If a process $p$ is in a {\sf
UDC($\alpha$)} state, it sends an $\alpha$-message repeatedly to all other
processes 
from which it has not received an acknowledgment,
telling them to perform $\alpha$.
Process $p$ performs $\alpha$ at time $m$ if, by time $m$, 
there is a set $S \subseteq \proc$ and $k \le |S|$ such that
(a) it is in a {\sf UDC($\alpha$)} state,
(b) its
failure detector has reported $\readfd{p}{S,k}$, (c) it has received
messages from all the processes in $\proc - S$ acknowledging $\alpha$, and
(d) $n - |S| > \min(t,n-1)-k$.  
Process $p$ continues to send
$\alpha$-messages to 
each
$q \in S$ until it either receives an
acknowledgment from $q$ or knows $q$ to be faulty.  
(Note that knowledge is only necessary for the protocol's termination.)
A process that receives
an $\alpha$-message from $p$ sends an acknowledgment to $p$ and goes into a
{\sf UDC($\alpha$)} state if it has not already done so.

To show that this protocol attains UDC, again it suffices to show that, in
every run, (1) if a process $p$ is in a {\sf UDC($\alpha$)} state, then $p$
will eventually perform $\alpha$ or crash and (2) if $p$ performs $\alpha$
then every other correct process performs $\alpha$.  
For (1), 
suppose that $p$ is in a {\sf UDC($\alpha$)} state in run $r$
and, by way of contradiction, $p$ neither performs $\alpha$ nor crashes.
Then $p$ repeatedly sends an $\alpha$-message in $r$ to every process
$q$. 
By R5, every correct process $q$ will get it infinitely often.
Since $q$ acknowledges $p$'s $\alpha$-message each time it gets it, by
R5, $p$ will eventually get an acknowledgment from every correct
process.  Since $p$ has a $t$-useful failure detector, if it is correct in
$r$, there will be a $t$-useful failure-detector event, say
$\readfd{p}{S,k}$, in $r_p(m)$ for some $m$.  Since $p$ eventually
gets acknowledgments from all the processes in $\proc - S$ (since these, at
least, are correct in $r$), it will eventually perform $\alpha$, according
to the algorithm.  Thus, (1) holds.

To see that (2) holds, the arguments for (1) show that if $p$ performs
$\alpha$ as a result of the failure-detector event $\readfd{p}{S,k}$, all
the processes in $\proc - S$ have received an $\alpha$ message (and hence
are in a {\sf UDC($\alpha$)} state) and $\proc - S$ contains at least one
correct process, say $q$, if there are any correct processes in $r$.  Since
$q$ continues to send $\alpha$-messages to all processes
from which it has not received an acknowledgment,
all the
correct processes in $r$ will eventually be in a {\sf UDC($\alpha$)} state.
It then follows from (1) that all the correct processes will perform
$\alpha$.  \eprf

\othm{UDCchar1} Suppose $\R$ is the system generated by a protocol that attains
UDC in a context with at most $t$ failures, $\R$ satisfies A1--A4 and
A5$_t$, and for each run $r \in \R$, 
if $F(r) \ne \proc$, then infinitely many actions are initiated in $r$. 
Then $\R^{f'}$ has $t$-useful  generalized failure detectors. \eothm
\medskip

\prf Again, it is easy to see that each 
correct
process $p$'s failure detector
satisfies generalized strong accuracy.  To show that it satisfies
generalized impermanent strong completeness, suppose that $p$ is correct.
Since infinitely many actions are initiated in $r$ (and hence also in
$f(r)$), there must be some
action $\alpha$ initiated by a correct process $p'$ in $f(r)$ at a time
after all the processes in $F(r)$ ($= F(f(r))$) have failed in $f(r)$.

Since $\R$ satisfies UDC, $p$ must eventually perform $\alpha$ in run $r$,
say at time $m$.  
As in the proof of Theorem~\ref{UDCchar}, using
Proposition~\ref{PropAccount}, we can conclude that
\begin{equation}\label{eq:contra3} 
(\R,r,m) \sat 
K_p \biggl(\bigvee_{q'\in \proc} \Box \neg \quitfla{q'} \rimp 
\bigvee_{q' \in \proc} \Bigl(K_{q'}
\initfla{p'}{\alpha} \land \Box \neg \quitfla{q'}\Bigr)\biggr).
\end{equation}

Suppose, by way of contradiction, that $p$ does not know that at least $k =
|F(r)| - n + t$ processes have crashed at $(r,m)$.  Then there must be a
point $(r^1,m') \sim_p (r,m)$ such that $k' < k$ processes have crashed by
$(r^1,m')$.  We must have $(\R,r,m) \sat \bigwedge_{q \in F(r)} \neg
K_q\initfla{p'}{\alpha}$, since
all the processes in $F(r)$ crashed in $r$ before $p'$ initiated $\alpha$.
Consequently, it follows that $(\R,r^1,m') \sat \bigwedge_{q\in 
F(r)} \neg K_p (K_q \initfla{p'}{\alpha})$.  
By repeated applications of
A4, there 
is 
a point $(r^2,m') \sim_p (r^1,m')$ such that $(\R,r^2,m') \sat
\bigwedge_{q \in F(r)} \neg K_q \initfla{p'}{\alpha}$.
(We are using the fact that $K_q \initfla{p'}{\alpha}$ is stable,
local to $q$, and insensitive to failures by $q$.  Thus,  if $\neg K_q
\initfla{p'}{\alpha}$ holds for some 
history of $q$, it holds for any prefix of that history
or a prefix followed by a $\quit{q}$ event.)
Thus, the only processes that may know $\initfla{p'}{\alpha}$
in $r$ are those in $\proc - F(r)$.  Since $|F(r)| = n-t+k$, we have that
$|\proc - F(r)| = t-k$.  Thus, at most $t-k$ processes know
$\initfla{p'}{\alpha}$ at the point $(r^2,m')$.

Let $F_1$ be the set of processes that have crashed by $(r^1,m')$ and let
$F_2$ be the set of processes that have crashed by $(r^2,m')$.  Since
$r^2_q(m')$ is a prefix of $r^1_q(m')$ for all $q \in \proc$, we must have
$F_2 \subseteq F_1$.  Recall that $|F_1| = k' < k$.  We now proceed much as
in the proof of Proposition~\ref{PropAccount} to construct a run extending
$(r^2,m')$ in which the processes in $F(r) - F_1$ do not crash and and do
not learn about $\initfla{p'}{\alpha}$. 

By A4, there exists a point $(r^3,m')$ such that $(r^3,m') \sim_{q}
(r^2,m')$ for $q \in F(r)$ and $(\R,r^3,m') \sat \neg \initfla{p'}{\alpha}$.
As in the previous application of A4, the set of processes that are faulty
at the point $(r^3,m')$ is a subset of $F_1$ and hence consists of at most
$k'$ processes.  By A1 and A5$_t$, there exists a run $r^4$ extending
$(r^3,m')$ such that $F(r^4) = (\proc - F(r)) \union F_1$.  Since $r^4$
extends $(r^3,m')$, we must have $r^4_q(m') = r^2_q(m')$ for all $q \in
F(r)$.  By A2, there exist runs $r^5$ and $r^6$ extending $r^2$ and $r^4$,
respectively, such that $r^5_q(m'') = r^6_q(m'')$ for $m'' \ge m'$.
Moreover, all the processes in $\proc-F(r) \union F_1$ crash by time $m' +
1$.  Clearly $p' \notin F(r)$ (since all the processes in $F(r)$ crash
before $p'$ initiates $\alpha$).  Thus, the event $\initfla{p'}{\alpha}$
does not appear in $r^6$.  It follows that $(\R,r^5,m'') \sat \bigwedge_{q
\in F(r)} \neg K_{q} \initfla{p'}{\alpha}$ for all $m'' \ge m'$ and $q \in
F(r)$, so $(\R,r^5,m') \models \bigwedge_{q \in F(r)}\Box \neg K_{q}
\initfla{p'}{\alpha} \land \bigwedge_{q \in (\proc - F(r)) \union F_1}
\Diamond \quitfla{q}$.  Since $(r,m) \sim_p (r^1,m')$, $(r^1,m') \sim_p
(r^2,m')$, and $r^5$ extends $(r^2,m')$, we must have $(r,m) \sim_p
(r^5,m')$.  But this gives us the desired contradiction to
(\ref{eq:contra3}).

Thus, $p$ must know about at least $k$ failures at the point $(r,m)$.  Let
$S$ be any set containing $F(r)$.  Our transformation from $\R$ to $\R^{f'}$
guarantees that eventually there will be a failure-detector event $(S,k)$
in $p$'s history, and this is a $t$-useful event.  \eprf

\paragraph{Acknowledgments:}  We thank Marcos Aguilera, Boris
Deianov, and Sam Toueg for their perceptive comments, particularly with
regard to assumption A4.
We especially thank Gil Neiger for his close reading of the paper and
many useful suggested changes.
Finally, we thank the reviewers of this paper for their detailed
and thoughtful comments.

\bibliographystyle{alpha}
\bibliography{z,joe}

\end{document}